\documentclass[apj]{emulateapj}
\usepackage{epsfig,psfig}

\begin{document}
\title{Discovery of a new photometric sub-class of faint and fast classical novae}

\author{M. M. Kasliwal\altaffilmark{1}, S. B. Cenko\altaffilmark{2},
S. R. Kulkarni\altaffilmark{1}, E. O. Ofek\altaffilmark{1},
R. Quimby\altaffilmark{1}, A. Rau\altaffilmark{3}}

\altaffiltext{1}{Astronomy Department, California Institute of Technology, 105-24, Pasadena, CA 91125, USA}
\altaffiltext{2}{Department of Astronomy, University of California at Berkeley, Berkeley, CA 94720, USA}
\altaffiltext{3}{Max-Planck Institut fuer Extraterrestrische Physik, 85748 Garching, Germany}
\email{mansi@astro.caltech.edu}

\begin{abstract}

We present photometric and spectroscopic follow-up of a sample of
extragalactic novae discovered by the Palomar 60-inch telescope during 
a search for ``Fast Transients In Nearest Galaxies'' (P60-FasTING). Designed
as a fast cadence (1-day) and deep ($g$\,$<$21\,mag) survey, P60-FasTING was particularly
sensitive to short-lived and faint optical transients. The P60-FasTING nova
sample includes 10 novae in M\,31, 6 in M\,81, 3 in M\,82, 1 in NGC\,2403 
and 1 in NGC\,891. This significantly expands the known sample of extragalactic
novae beyond the Local Group, including the first discoveries in a starburst environment. 
Surprisingly, our photometry shows that this sample is quite inconsistent with the 
canonical Maximum Magnitude Rate of Decline (MMRD) relation for classical novae. 
Furthermore, the spectra of the P60-FasTING sample are indistinguishable from classical novae. 
We suggest that we have uncovered a 
sub-class of faint and fast classical novae in a new phase space in luminosity-timescale of
optical transients. Thus, novae span two orders of magnitude in both luminosity and time. Perhaps, the MMRD, which 
is characterized only by the white dwarf mass, was an over-simplification. Nova physics appears to be characterized
by quite a rich four-dimensional parameter space in white dwarf mass, temperature, composition and accretion rate.
\end{abstract}


\section{Introduction}
Since the discovery of classical novae, astronomers have pursued their use as standard candles 
to determine distances (see \citealt{h29}). \citet{z36} first noticed some regularity in nova light
curves and termed this the ``life-luminosity'' relation. \citet{a56} undertook a comprehensive 
search for novae in M31, discovering thirty novae in 290 nights, and found a clear relation 
--- luminous novae evolve faster than less luminous novae. The modern name for this observation
is the maximum-magnitude rate-of-decline relation (MMRD relation). 

The MMRD relation has attracted considerable theoretical attention (e.g. \citealt{l92}).
The basic idea is that the relation is entirely due to the mass of the accreting white dwarf. 
The more massive the white dwarf, the higher the surface gravity, the higher 
the pressure at the base of envelope and stronger the thermonuclear runaway (and hence, higher the peak luminosity).
Also, the more massive the white dwarf, the smaller the envelope mass to attain the critical  
pressure for thermonuclear runaway (TNR) and hence, faster the decline. 

In more recent times, \citet{dl95} used a sample of novae in M31 and LMC to propose
an arctangent relation between the peak luminosity and rate of decline. \citet{dd00} used a sample
of Galactic novae to propose a linear relation between the same two parameters. \citet{dbk+06} used a score
of novae in M31 from POINT AGAPE survey and claimed their observations were consistent with 
the \citet{dl95} formulation of the MMRD. 

In comparison to supernovae, classical novae are not very luminous. Hence, searches have 
traditionally focussed only on the Milky Way and its nearest neighbors 
(Andromeda and the Large Magellanic Cloud). 
\citet{hsg+08} looked into archival data and found 49 nova candidates{\footnote{We use the term candidate where 
the light curve is very sparse and/or there is no spectrosopic confirmation}} in M\,81 
in the past 20 years --- unfortunately, these candidates neither have light curves nor spectra. 
\citet{fcj03} undertook a search for novae using 
24 orbits of the Hubble Space Telescope and found nine nova candidates in M49. Even with their
sparsely sampled light curves for nine novae, they concluded that novae are not
good standard candles. Another survey, CFHT-COVET{\footnote{{\bf C}anada {\bf F}rance {\bf H}awaii {\bf T}elescope {\bf CO}ma {\bf V}irgo {\bf E}xploration for {\bf T}ransients}} 
(aimed at finding transients in the gap between novae and supernovae) 
found a dozen nova candidates in many galaxies in the Virgo supercluster, 
including some in the far outskirts of galaxies (Kasliwal et al 2010, in prep).

Here, we report on novae discovered in high cadence 
monitoring observations of a representative collection of galaxies
with distance less than that of the Virgo cluster. The original motivation
of this search -- P60-FasTING{\footnote{{\bf P}alomar {\bf 60}-inch 
{\bf Fas}t  {\bf T}ransient {\bf I}n {\bf N}earest {\bf G}alaxies}}
was to explore rapid transients (those which last less than a couple of nights)
in the nearest galaxies. A strong spectroscopic follow-up effort was a part of P60-FasTING.
The survey was capable of finding novae in the major galaxies out
to 4\,Mpc: M\,31, M\,81, the star-burst M\,82 and 
NGC\,2403. We present our sample of 21 transients, which although spectroscopically
indistinguishable from classical novae, photometrically occupy a new region of phase space. 

The paper is organized as follows: \S\,\ref{sec:obs} describes the discovery, photometric
and spectroscopic follow-up observations of this nova sample, \S\,\ref{sec:analysis} describes
the data analysis, \S\,\ref{sec:discussion} discusses the implications and \S\,\ref{sec:conclusion}
presents our conclusion.

\begin{table*}[!hbt]
\begin{center}
\caption[]{\bf Novae Discovered by P60-FasTING}
\begin{tabular}{lllllllll}
\hline
\hline
Nova & Host & Discovery Date &  RA(J2000) & DEC(J2000) & Offset from Host & Reference\cr 
\hline 
P60-NGC2403-090314 & NGC2403 &  2009 Mar 14.160 & 07:36:35.00 & +65:40:20.8 & 101.0"W, 252.0"N & \cite{k+09c} \cr 
P60-M82OT-090314 & NGC3034 & 2009 Mar 14.496 & 09:56:12.60 &  +69:41:32.3 & 104.2"E, 48.2"N  & \nodata \cr  
P60-M81OT-090213 & NGC3031 & 2009 Feb 13.404 & 09:55:35.96 & +69:01:51.0 & 15"E, 124"S &  \cite{k+09b}  \cr
P60-M31OT-081230 (2008-12b) & NGC224 & 2008 Dec 30.207 &  00:43:05.03 & +41:17:52.3 & 233.4"E,103.8"N &  \cite{k+09a} \cr
P60-M81OT-081229 & NGC3031 & 2008 Dec 29.373 &  09:55:38.15 & +69:01:43.6 & 26.7"E, 131.4"S & \cite{rks09} \cr
P60-M81OT-081203 & NGC3031 & 2008 Dec 3.303 & 09:55:16.92 & +69:02:17.7 & 87.2"W, 97.4"S & \cite{k+08f} \cr
P60-M82OT-081119 & NGC3034 & 2008 Nov 19.536 & 09:55:58.39 &  +69:40:56.2 & 29.5"E, 10.4"N &\cite{k+08e} \cr 
P60-M81OT-081027 & NGC3031 & 2008 Oct 27.402 & 09:55:36.11 & +69:03:22.0 & 15.8"E, 33.1"S & \cite{k+08d} \cr
P60-M81OT-080925 & NGC3031 & 2008 Sep 25.49 & 09:55:59.35 & +69:05:57.1 & 2.35'E, 2.03'N & \cite{k+08c} \cr
P60-M31OT-080915 (2008-09c) & NGC224 & 2008 Sep 15.36 & 00:42:51.42 & +41:01:54.0 & 1.34'E, 14.24'S & \cite{k+08h} \cr 
P60-M31OT-080913 (2008-09a) & NGC224 & 2008 Sep 13.18 & 00:41:46.72 & +41:07:52.1 & 10.8'W, 8.3'S & \cite{k+08g} \cr
P60-NGC891OT-080813 & NGC891 & 2008 Aug 13.45 & 02:22:32.70 & +42:21:56.1 & 8"W,59"N &  \cite{k+08b}\cr
P60-M31OT-080723 (2008-07b) & NGC224 & 2008 Jul 23.33 & 00:43:27.28 & +41:10:03.3 & 8.1'E,6.1'S & \cite{k+08i}\cr
P60-M82OT-080429 & NGC3034 & 2008 Apr 29.24 & 09:55:21.00 & +69:39:42.0 &  165" W, 64" S & \cite{k+08a}\cr
P60-M81OT-071213 & NGC3031 & 2007 Dec 13.40 & 09:55:25.98 & +69:04:34.8 & 40"W,40"N & \cite{k+07} \cr
\hline
\hline
\label{tab:allnovae}
\end{tabular}
\end{center}

\end{table*}

\begin{table*}[!hbt]
\begin{center}
\caption[]{\bf Additional M31 Novae}
\begin{tabular}{llll}
\hline
\hline
Nova Name & Discovery Date & Classification & Reference\cr  
\hline 
2007-10a & 54380.606 & Fe II & \cite{pbs+07,gr07} \cr 
2007-11f & 54433.716 & \nodata & \cite{ovk+07} \cr 
2007-12b & 54444.528 & He/N & Nakamo,Hornoch \cite{lrr+07,bds+09} \cr 
2008-08c & 54708.127  & \nodata & \cite{vol+08};Hornoch \cr 
2008-10b & 54759.698  & Fe II & \cite{hpb+08,dco+08,bfh+08} \cr 
2008-11a & 54774.438  & Hybrid & Nishiyama;Hornoch \cite{scb+08} \cr 
\hline
\hline
\label{tab:addm31novae}
\end{tabular}
\end{center}
\end{table*}

\section{Observations}
\label{sec:obs}
\subsection{Experiment Design}
P60-FasTING was designed with the specific goal of probing new phase space, 
particularly, fast transients with peak luminosity in the gap between novae
and supernovae. The sample of galaxies included the brightest and nearest
galaxies ($<$20\,Mpc, majority around 10\,Mpc). 
The survey was undertaken in a single filter (primarily Gunn-$g$ and some Gunn-$i$ 
data just around full moon). The limiting magnitude was typically Gunn-$g<$21 and cadence was $<$1\,day.  
The field of view 
of P60 is 13.5$^{\prime}\times$13.5$^{\prime}$ and all galaxies except M31 were covered 
in a single pointing. For M31, five pointings were chosen to cover a larger fraction of the galaxy.  

A real-time data reduction and transient search pipeline was written and implemented
in April 2008. P60-FasTING ended in March 2009. 
The search pipeline was written in \texttt{python}. 
A deep reference image was constructed by combining images from several of the best seeing dark nights. 
Next, \texttt{wcsremap} was used to align every new image with the reference and \texttt{hotpants} 
was used to compute a convolution kernel prior to image subtraction (both codes supplied by A. Becker 
{\footnote {http://www.astro.washington.edu/users/becker/c\_software.html}}).
Although the image subtraction software was quite sophisticated in its convolution of the 
new image to  match the reference prior to subtraction, we suffered from a large number of 
false positives.
To distill the false subtraction residuals from the bonafide astrophysical
sources, a variety of automatic filters were used (e.g. the shape characteristics of the PSF of the candidate,
how well does it resemble the PSF characteristics of other stars in the image). However, the final 
step in the vetting process was done by human eyes on candidate thumbnails every morning.
Due to the myriad trade-offs for maximum completeness and minimum contamination, the
complex issue of quantifying the completeness of the nova sample is beyond the scope of this
publication.

Our survey was sensitive to classical novae only in a handful of the nearest galaxies in our
sample (distance, d$<$4\,Mpc). Classical novae discovered by P60-FasTING are summarized in 
Table~\ref{tab:allnovae}. Some novae in M31 were announced by different groups before 
P60-FasTING's first detection (usually due to bad weather at Palomar) --- these are 
summarized in Table~\ref{tab:addm31novae}.

\begin{figure}[!hbt]
\begin{center}
\epsfig{figure=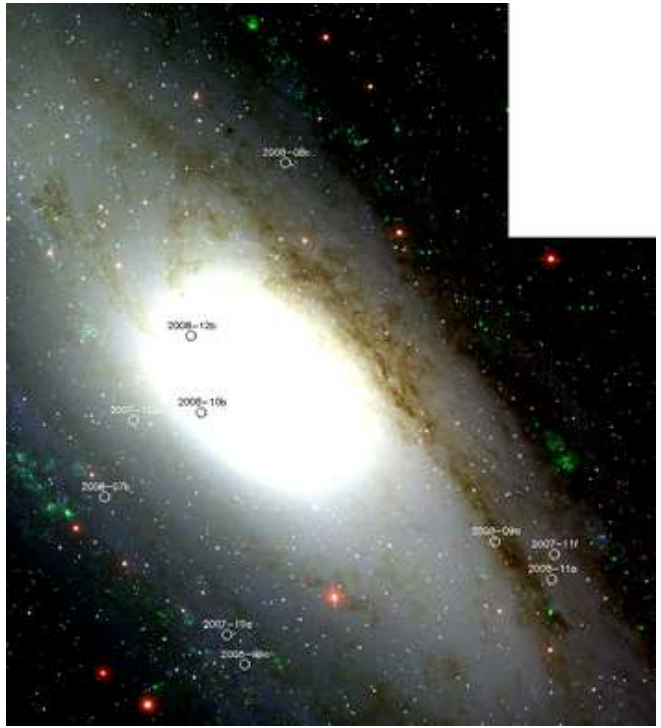, width=\linewidth, angle=0}
\caption{\small Location of ten novae in M31. Background image is a mosaic based on Massey data.
\label{fig:mosm31}}
\end{center}
\end{figure}

\begin{figure}[!hbt]
\begin{center}
\epsfig{figure=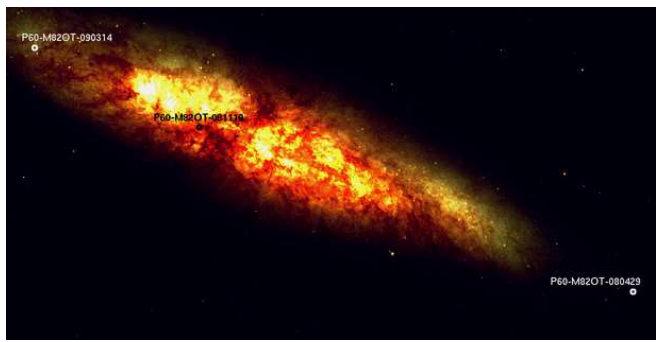, width=\linewidth, angle=0}
\caption{\small Location of three novae discovered by P60-FasTING in the starburst environment of M82. 
Background image is an HST/ACS mosaic. 
\label{fig:mosm81}}
\end{center}
\end{figure}

\begin{figure}[!hbt]
\begin{center}
\epsfig{figure=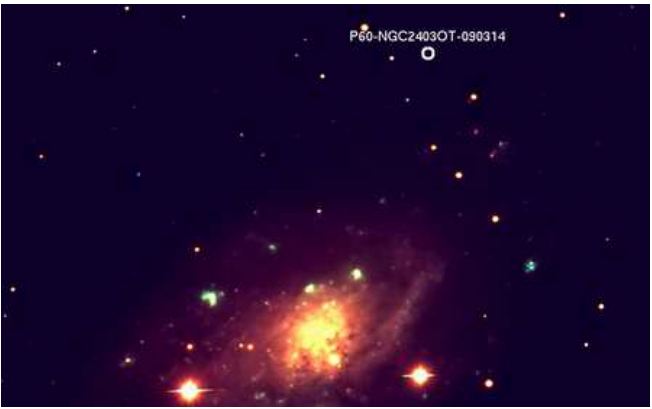, width=\linewidth, angle=0}
\caption{\small Location of one nova discovered by P60-FasTING in NGC2403. Background image is a deep co-add of P60 data.
\label{fig:mos2403}}
\end{center}
\end{figure}

\begin{figure}[!hbt]
\begin{center}
\epsfig{figure=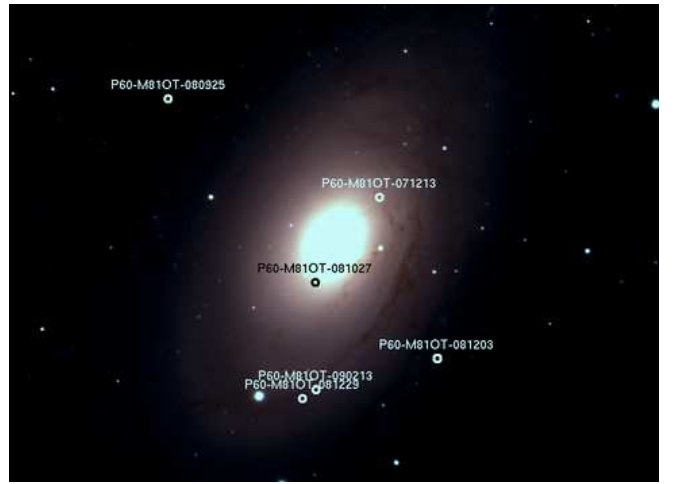, width=\linewidth, angle=0}
\caption{\small Location of six novae discovered by P60-FasTING in M81. Background image is a color mosaic using SDSS data.
\label{fig:mosm81}}
\end{center}
\end{figure}

\subsection{Photometry}
The robotic Palomar 60-inch has a standard data reduction pipeline (Cenko et al. 2006). This
pipeline performs basic detrending (flat-fielding and bias subtraction) and
computes an astrometric solution. In August 2008, we added a new functionality:
computation of a photometric solution. We used the SDSS catalog where available,
otherwise the NOMAD catalog. Note that where NOMAD was used (e.g. M31), the transformation 
from Johnson $UBVRI$ magnitudes to SDSS $ugriz$ magnitudes was done following \citet{jga06}.

To compute a light curve, we first measured the magnitude of the nova on the subtracted image.
The subtracted image was scaled to the same flux level as the new image. Thus,
we measured the magnitude of $\sim$150 reference stars on the new image
to compute a relative zeropoint with appropriate outlier rejection. Finally,
we applied this relative zeropoint to the instrumental magnitude of the nova.


\begin{figure*}[!hbt]
\begin{center}
\begin{tabular}{cc}
\psfig{figure=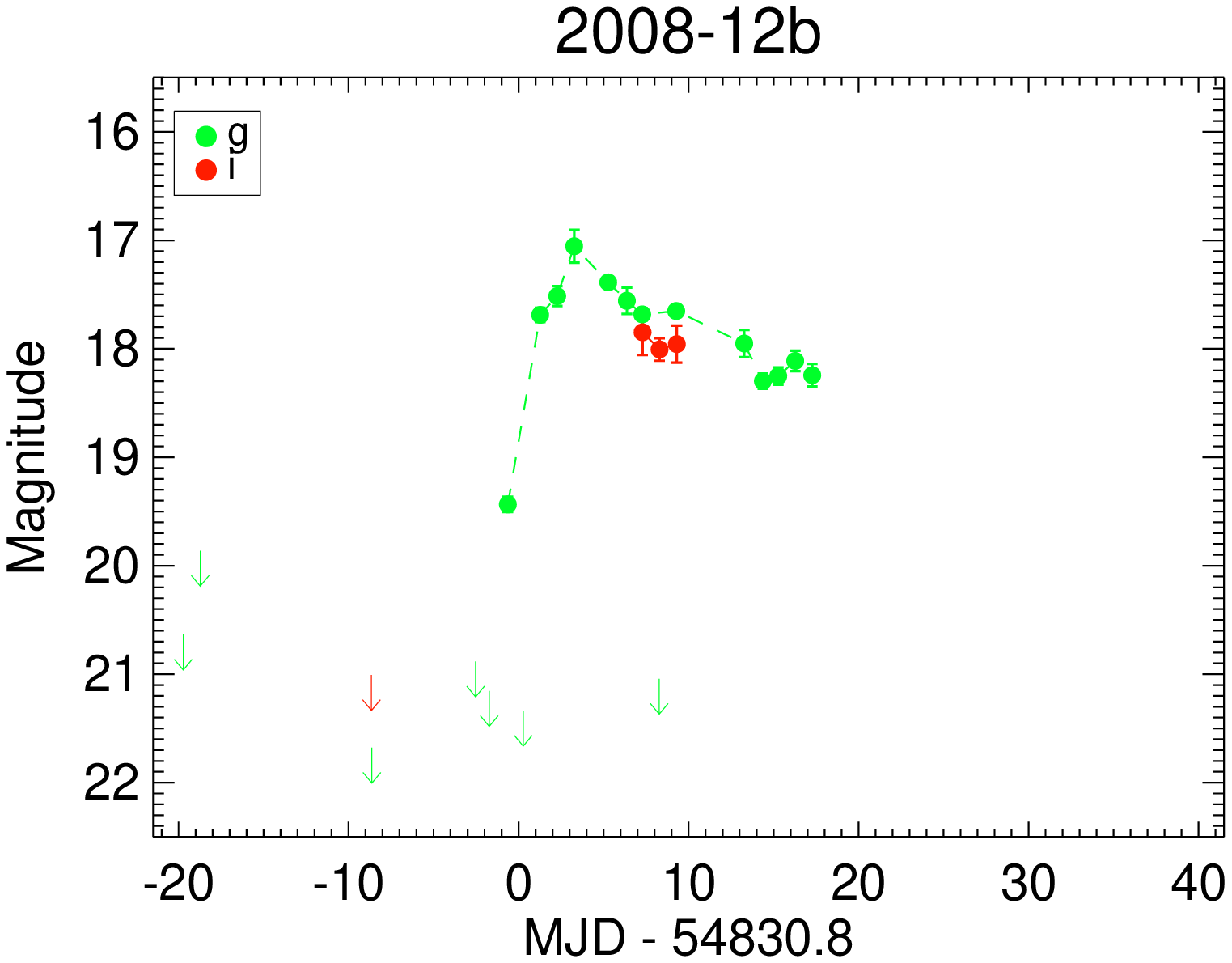,width=0.48\linewidth, angle=0} &
\psfig{figure=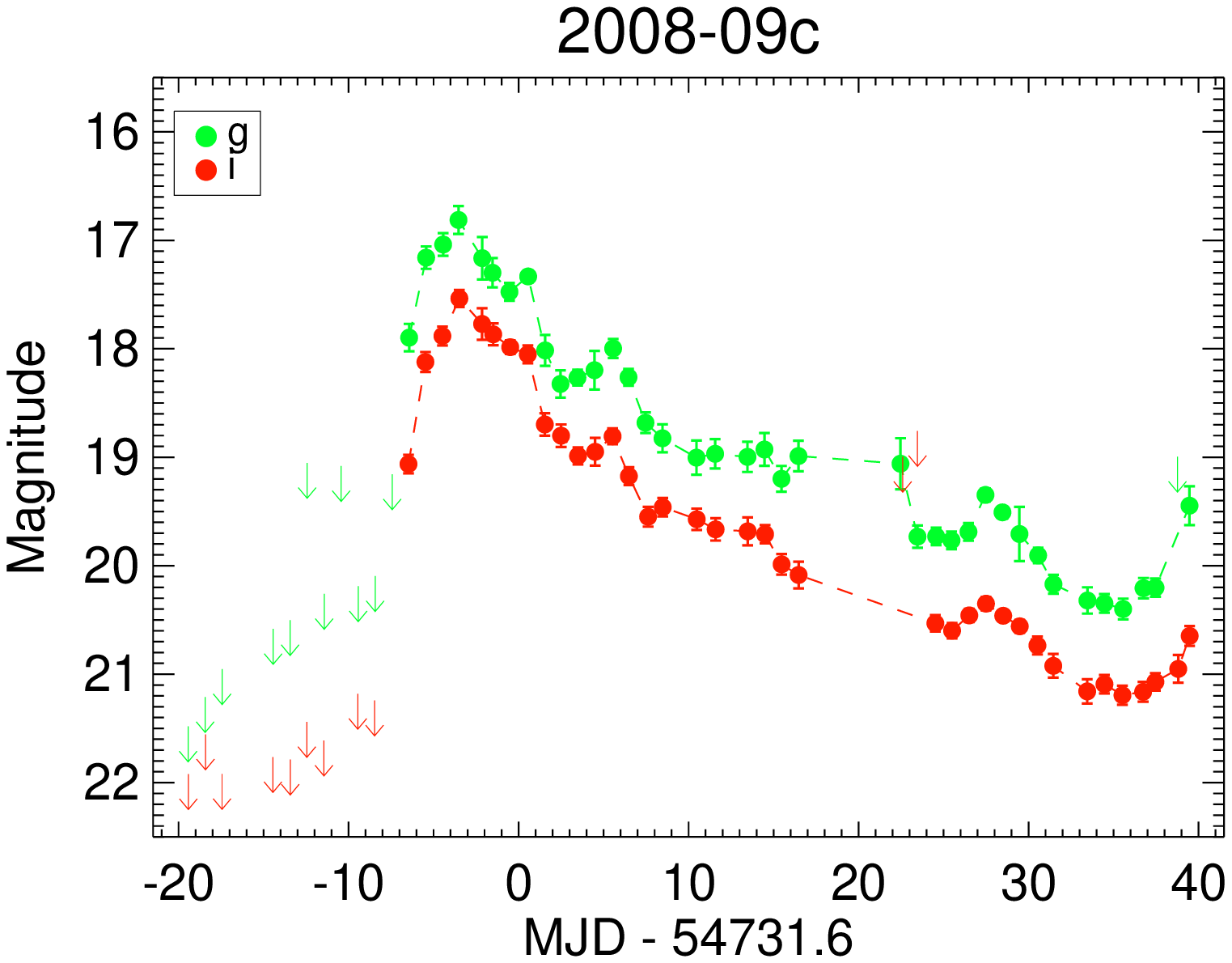,width=0.48\linewidth, angle=0} \\
\psfig{figure=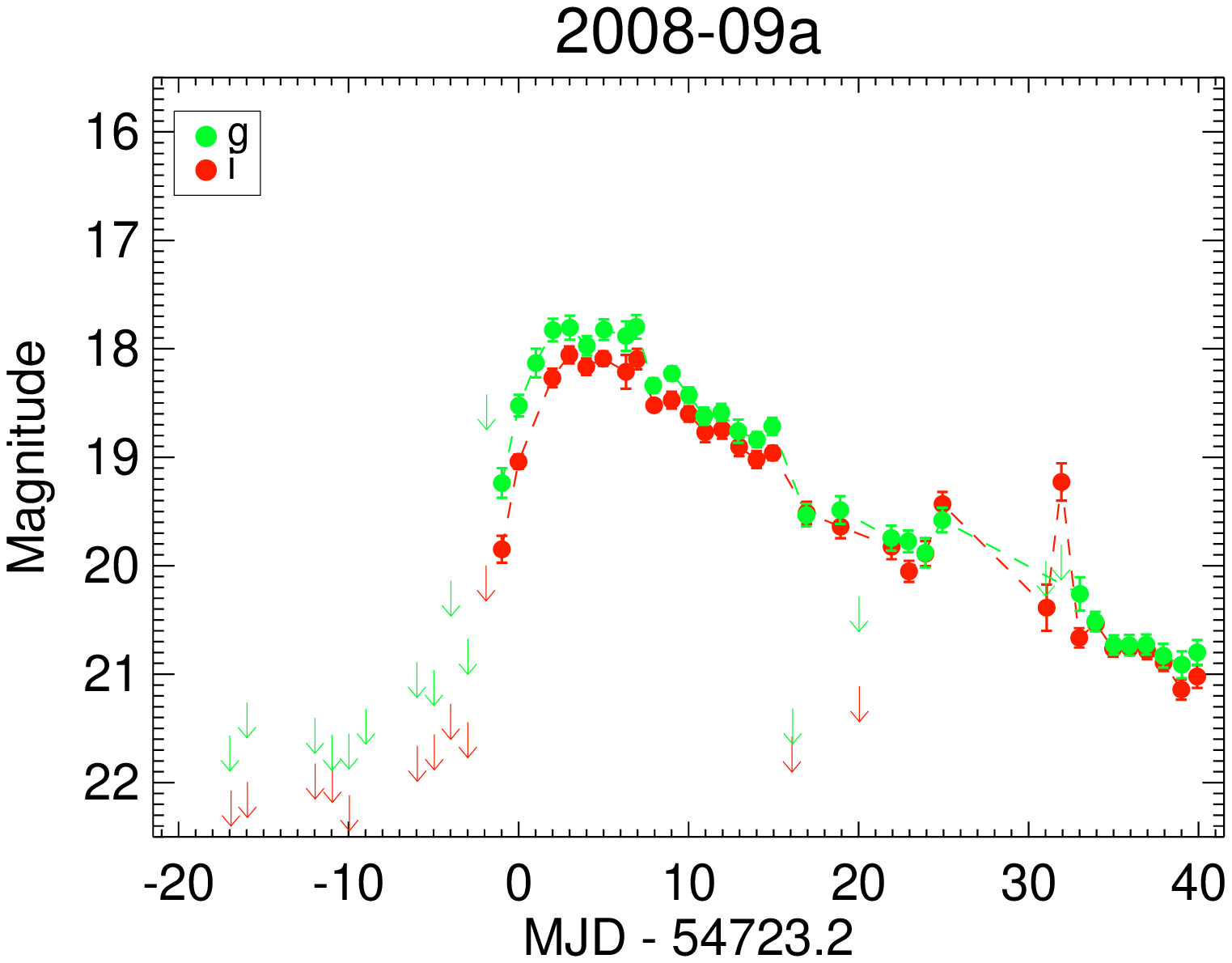,width=0.48\linewidth, angle=0} &
\psfig{figure=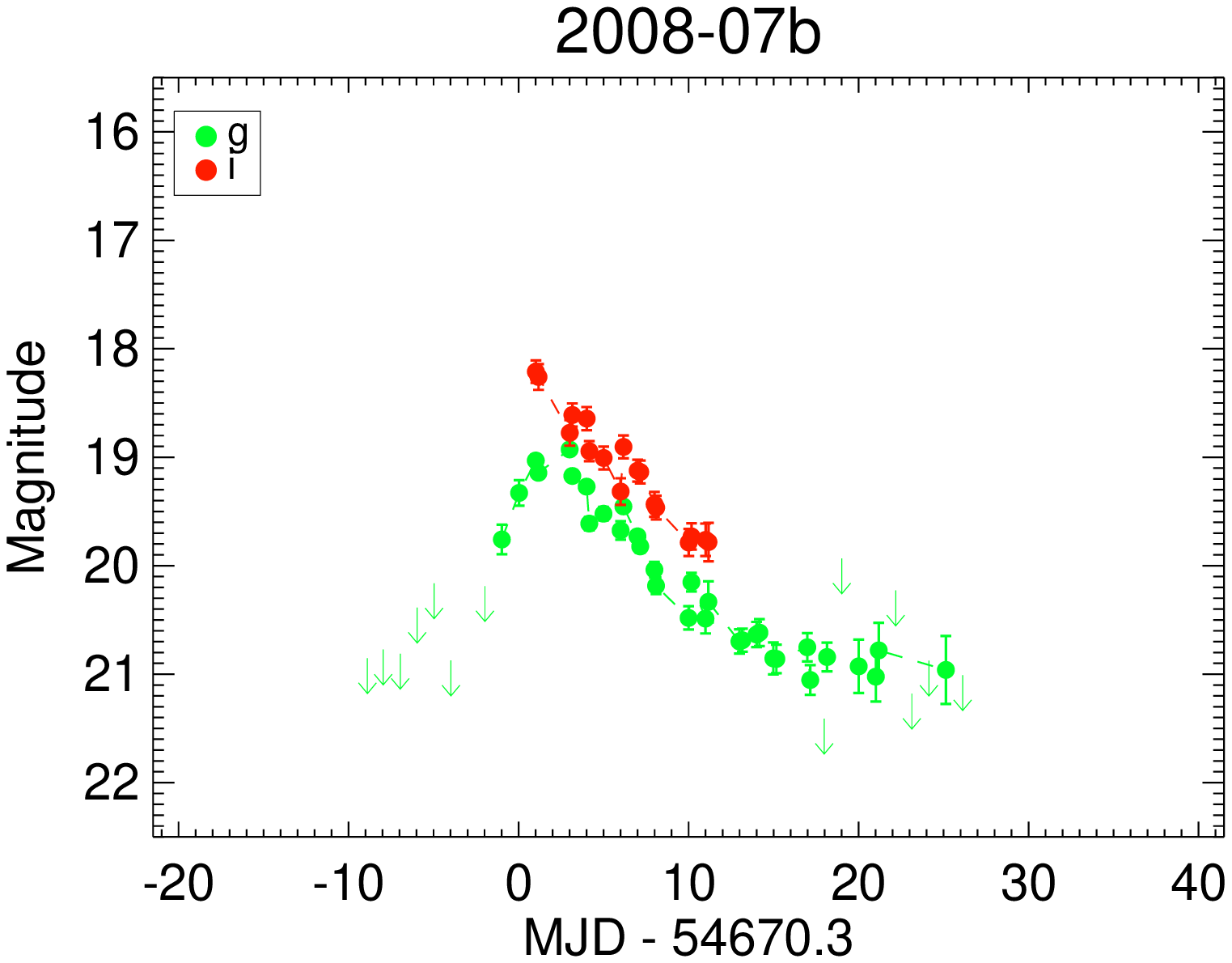,width=0.48\linewidth, angle=0} \\
\end{tabular}
\caption{\small Lightcurves of novae in M31 discovered by P60-FasTING. Note the well-sampled rise. 
\label{fig:m31lc}}
\end{center}
\end{figure*}

\begin{figure*}[!hbt]
\begin{center}
\begin{tabular}{cc}
\psfig{figure=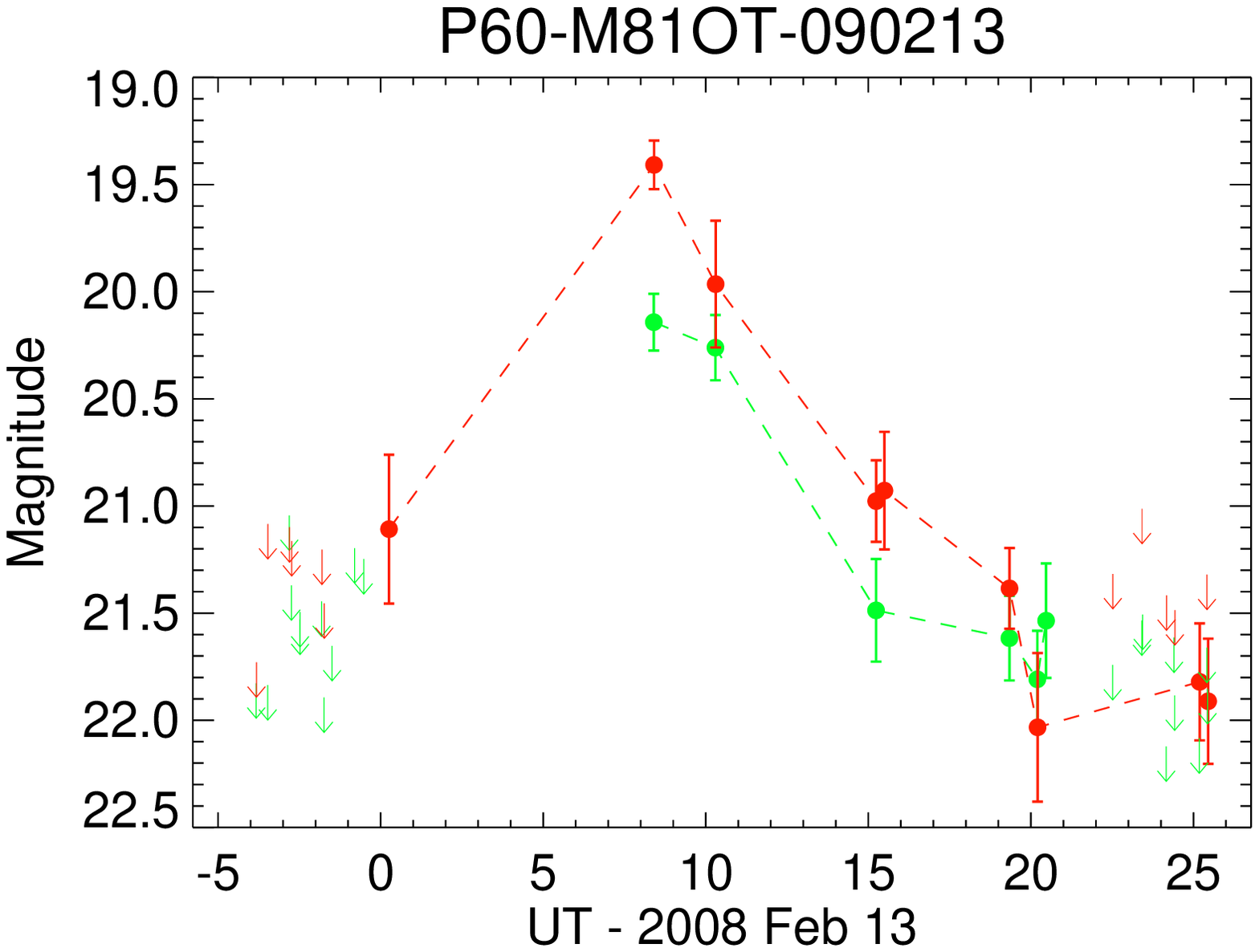,width=0.48\linewidth, angle=0} &
\psfig{figure=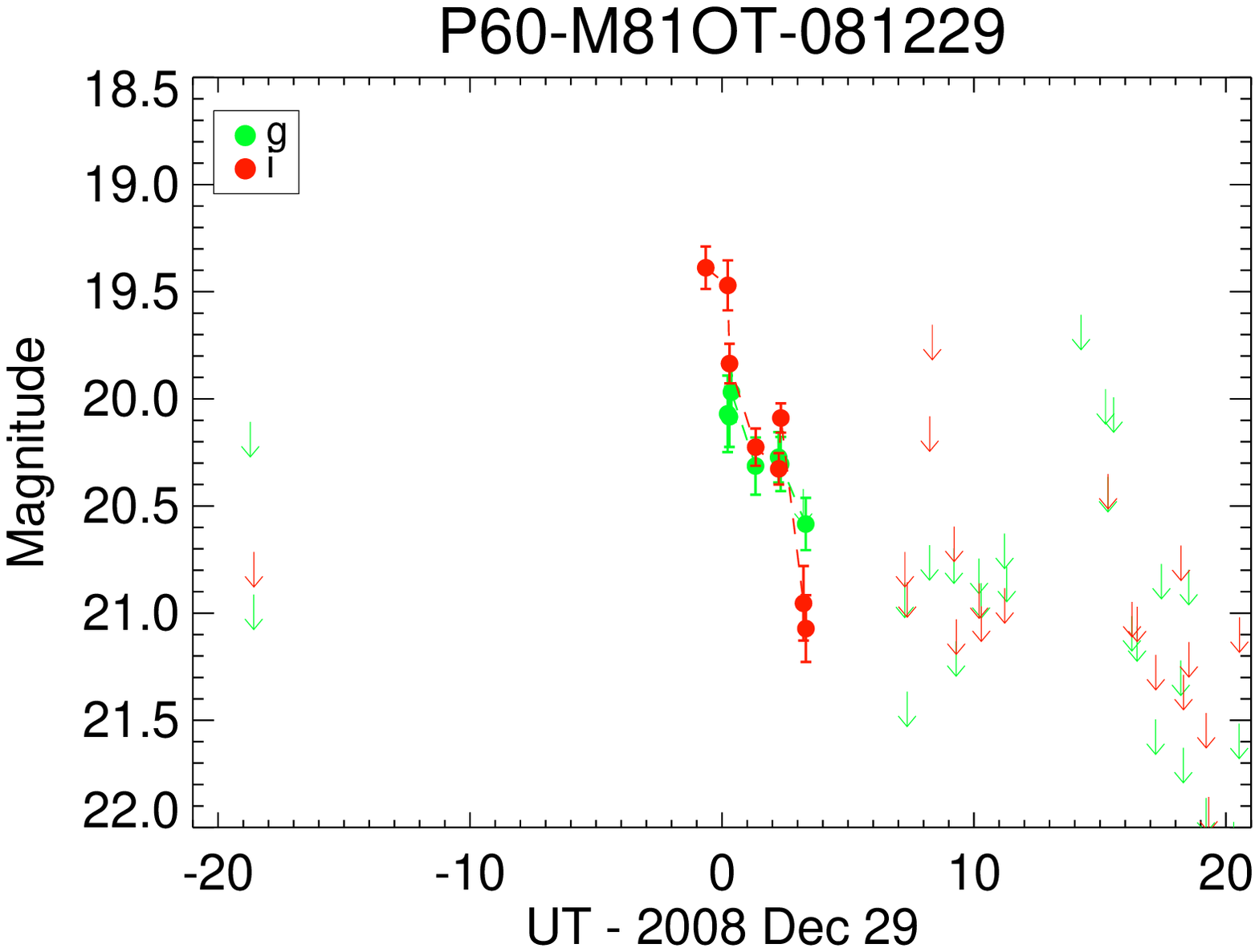,width=0.48\linewidth, angle=0} \\ 
\psfig{figure=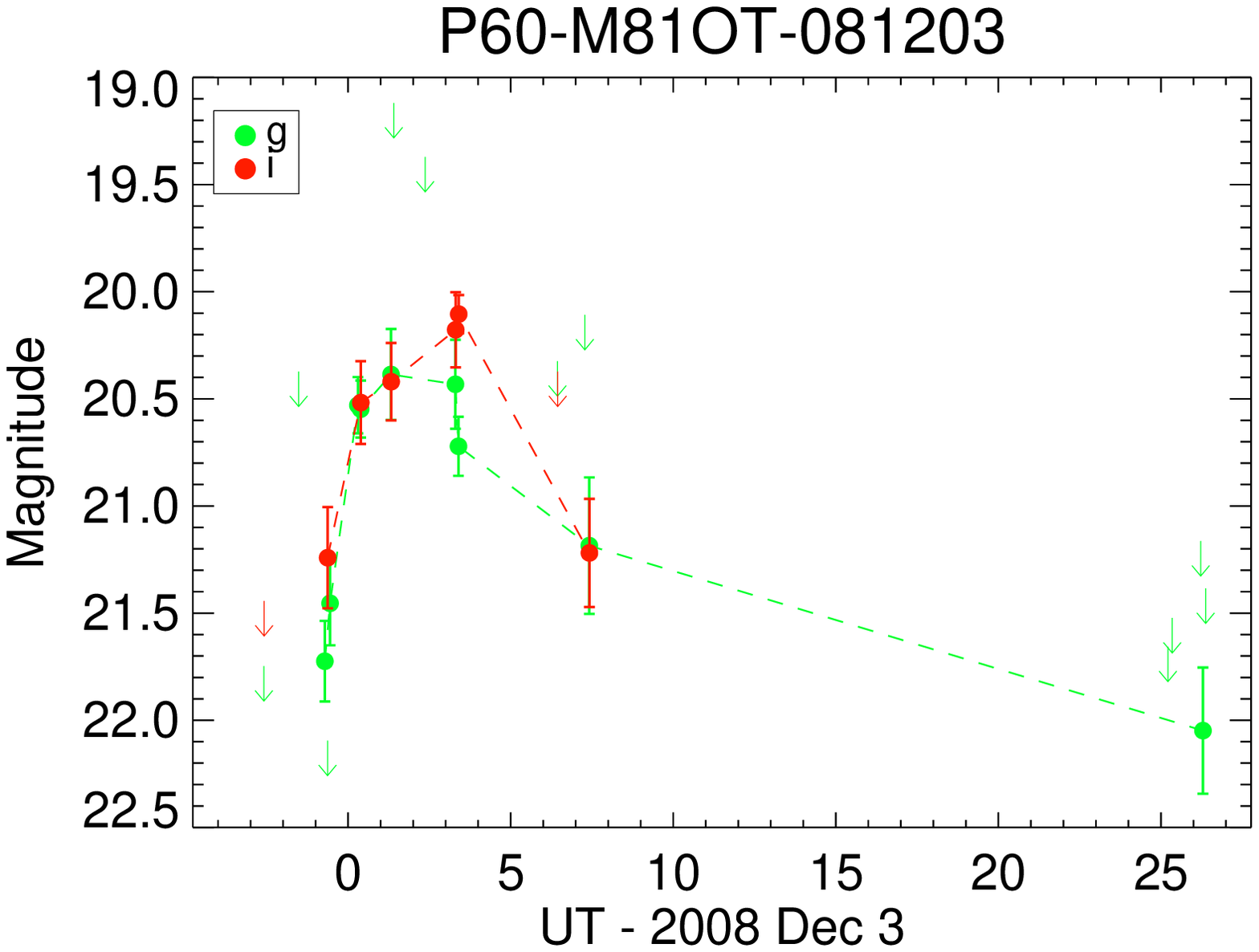,width=0.48\linewidth, angle=0} &
\psfig{figure=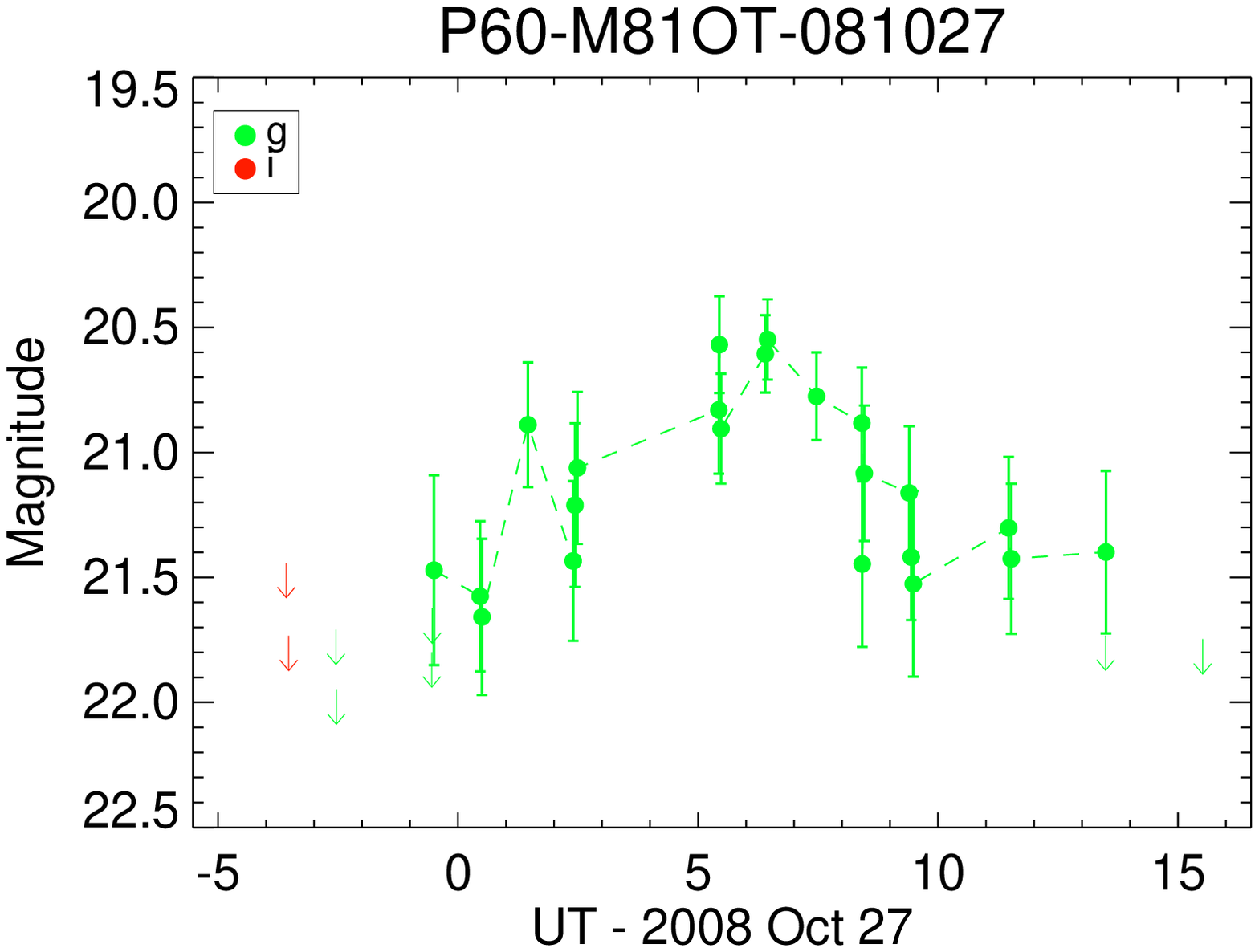,width=0.48\linewidth, angle=0} \\
\psfig{figure=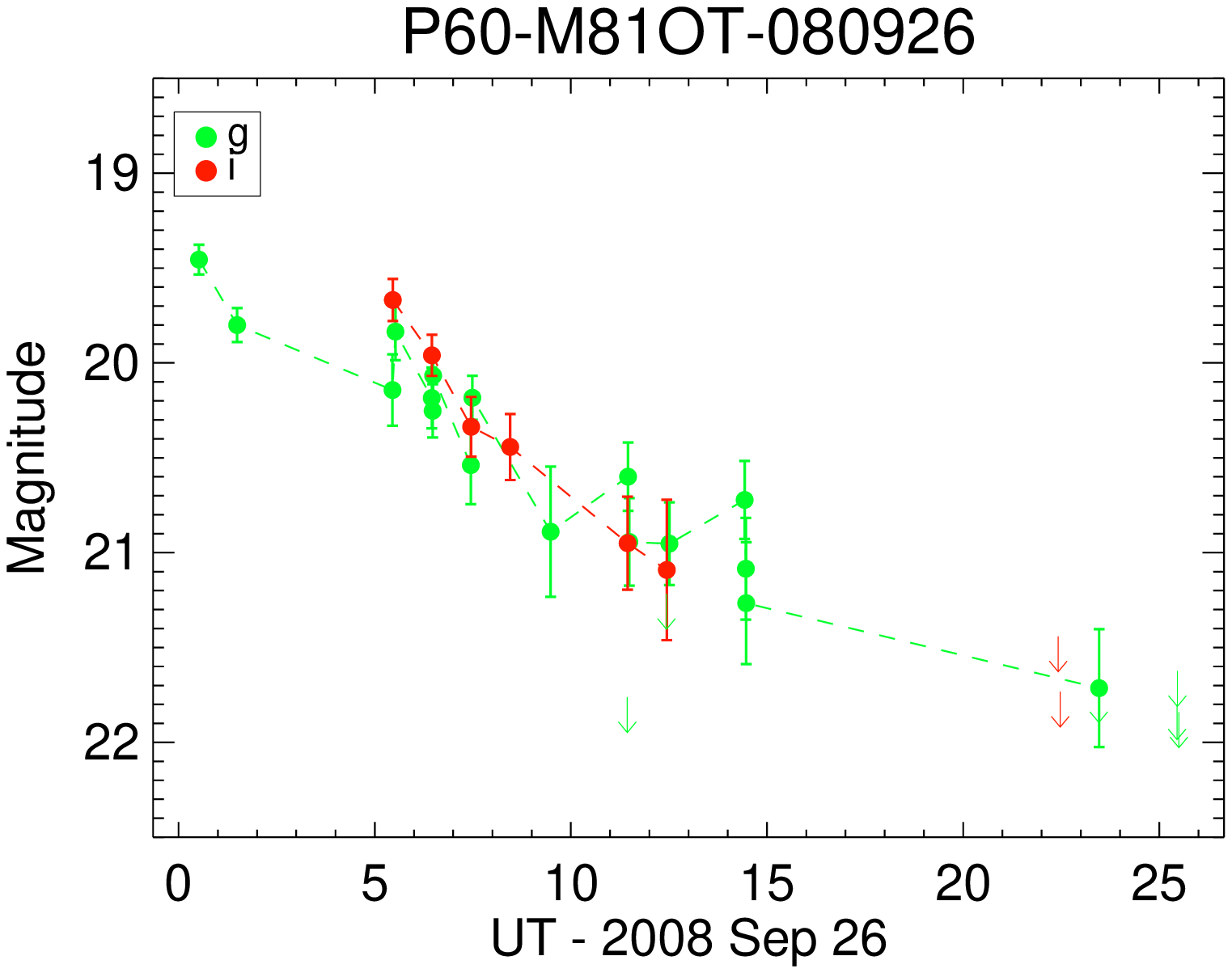,width=0.48\linewidth, angle=0} &
\psfig{figure=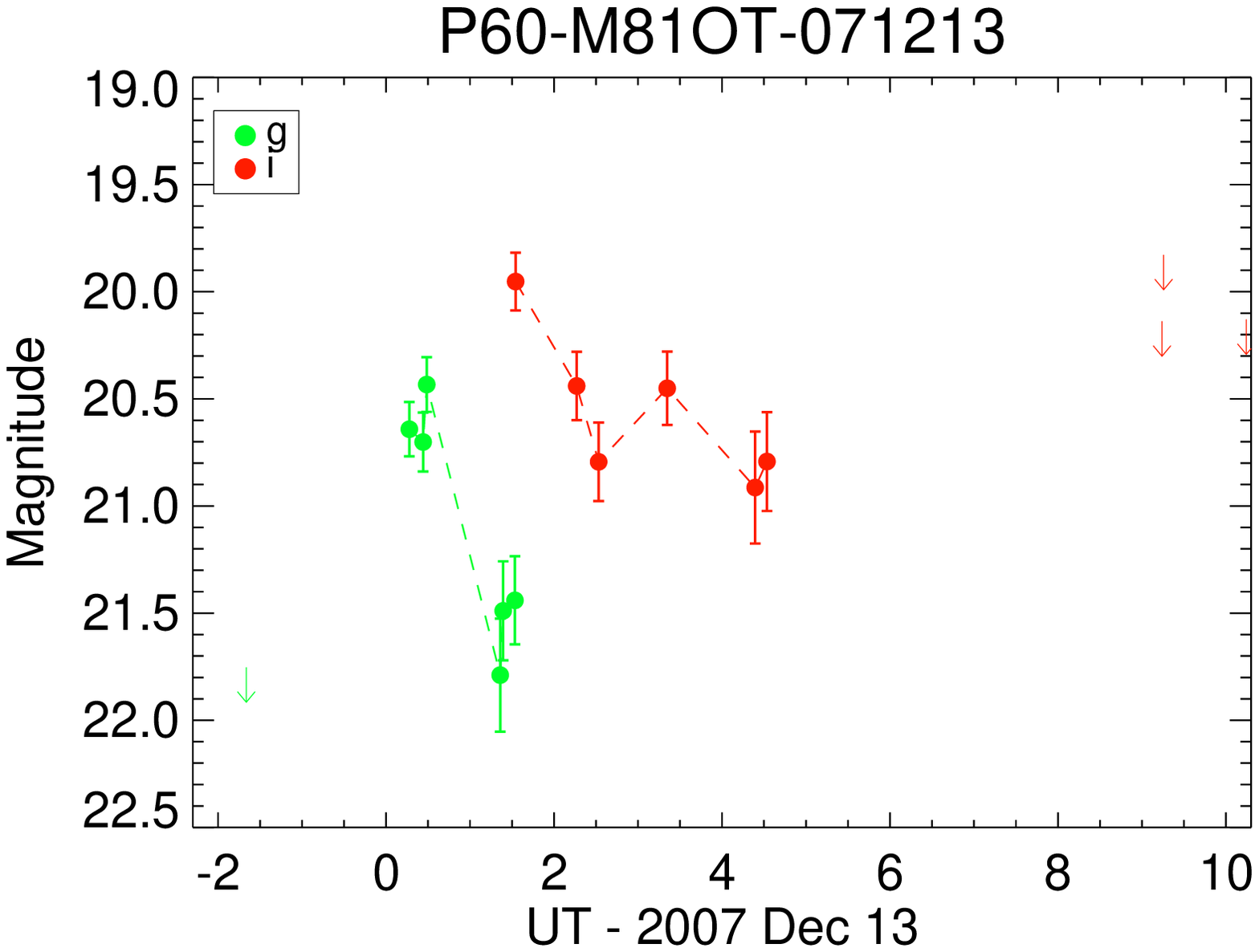,width=0.48\linewidth, angle=0} \\
\end{tabular}
\caption{\small Lightcurves of novae in M81. 
\label{fig:m81lc}}
\end{center}
\end{figure*}

\begin{figure*}[!hbt]
\begin{center}
\begin{tabular}{cc}
\psfig{figure=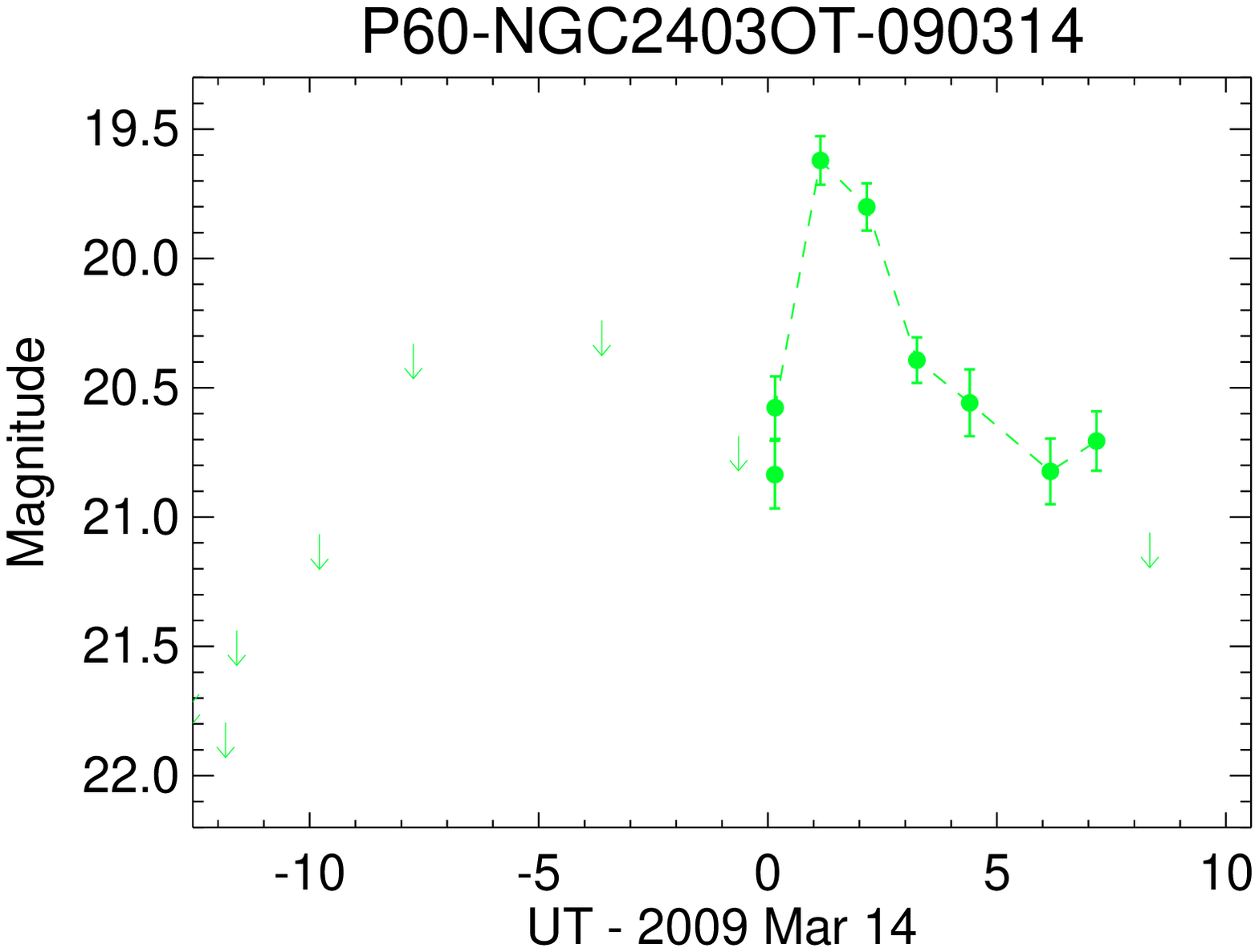,width=0.48\linewidth, angle=0} &
\psfig{figure=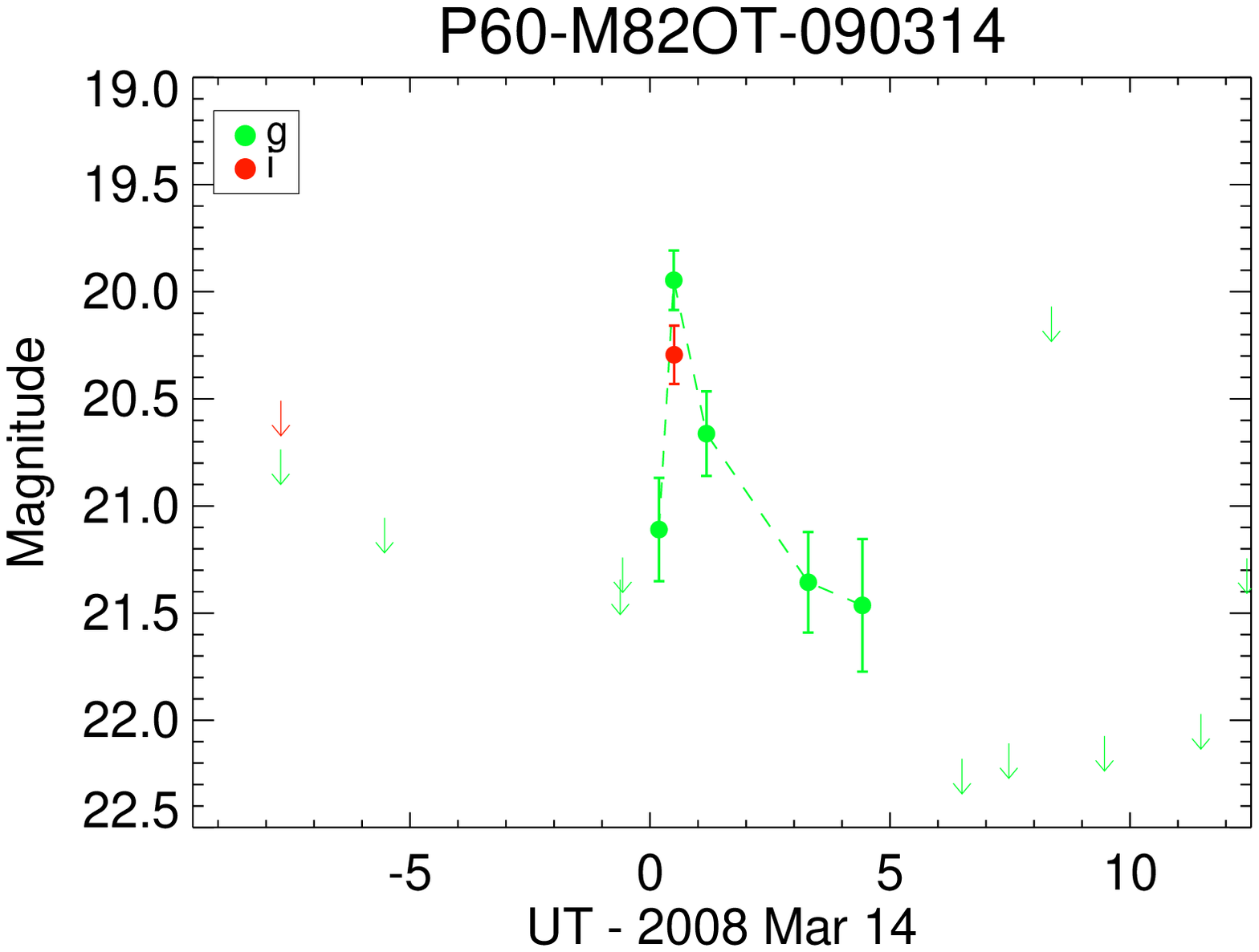,width=0.48\linewidth, angle=0} \\
\psfig{figure=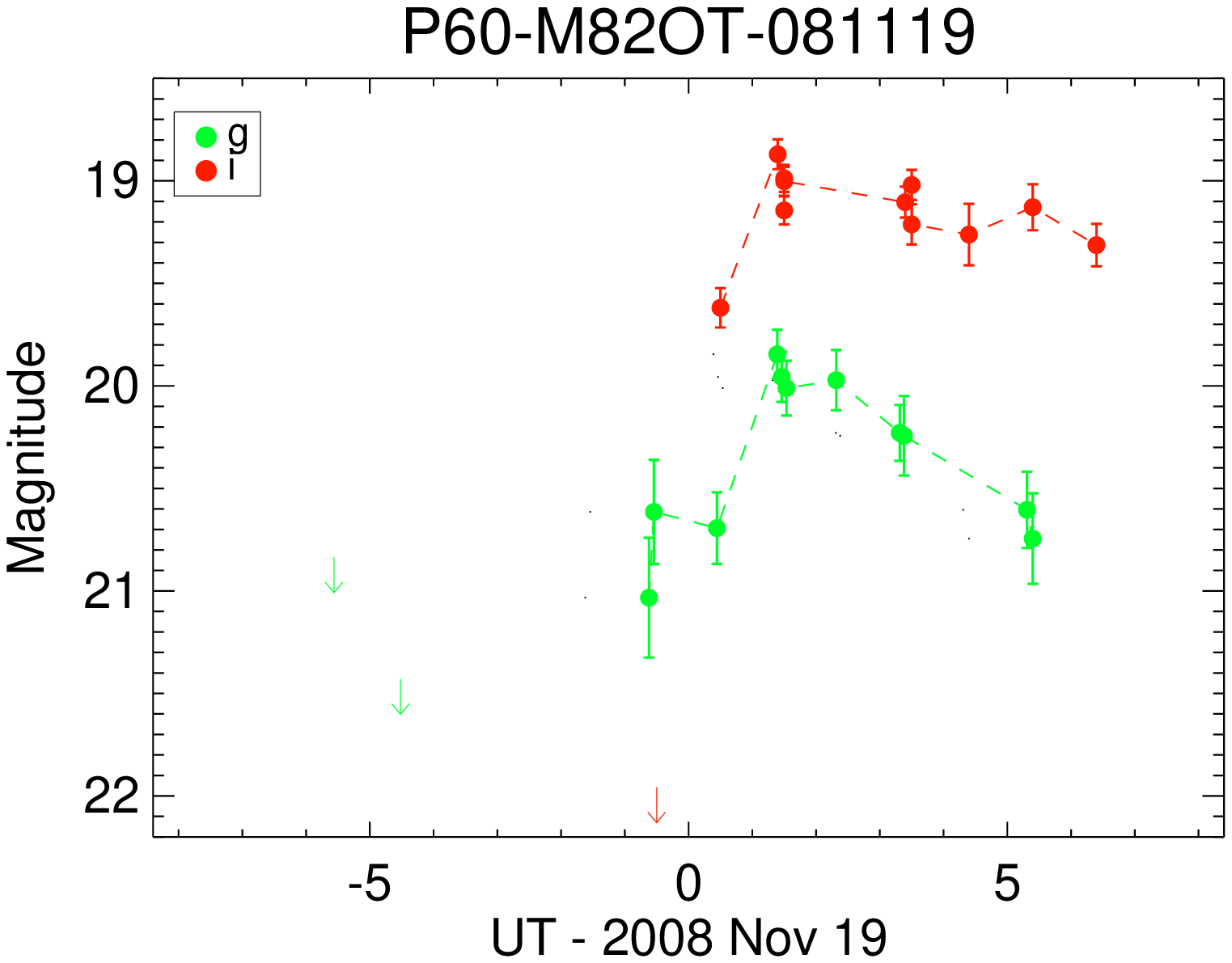,width=0.48\linewidth, angle=0} &
\psfig{figure=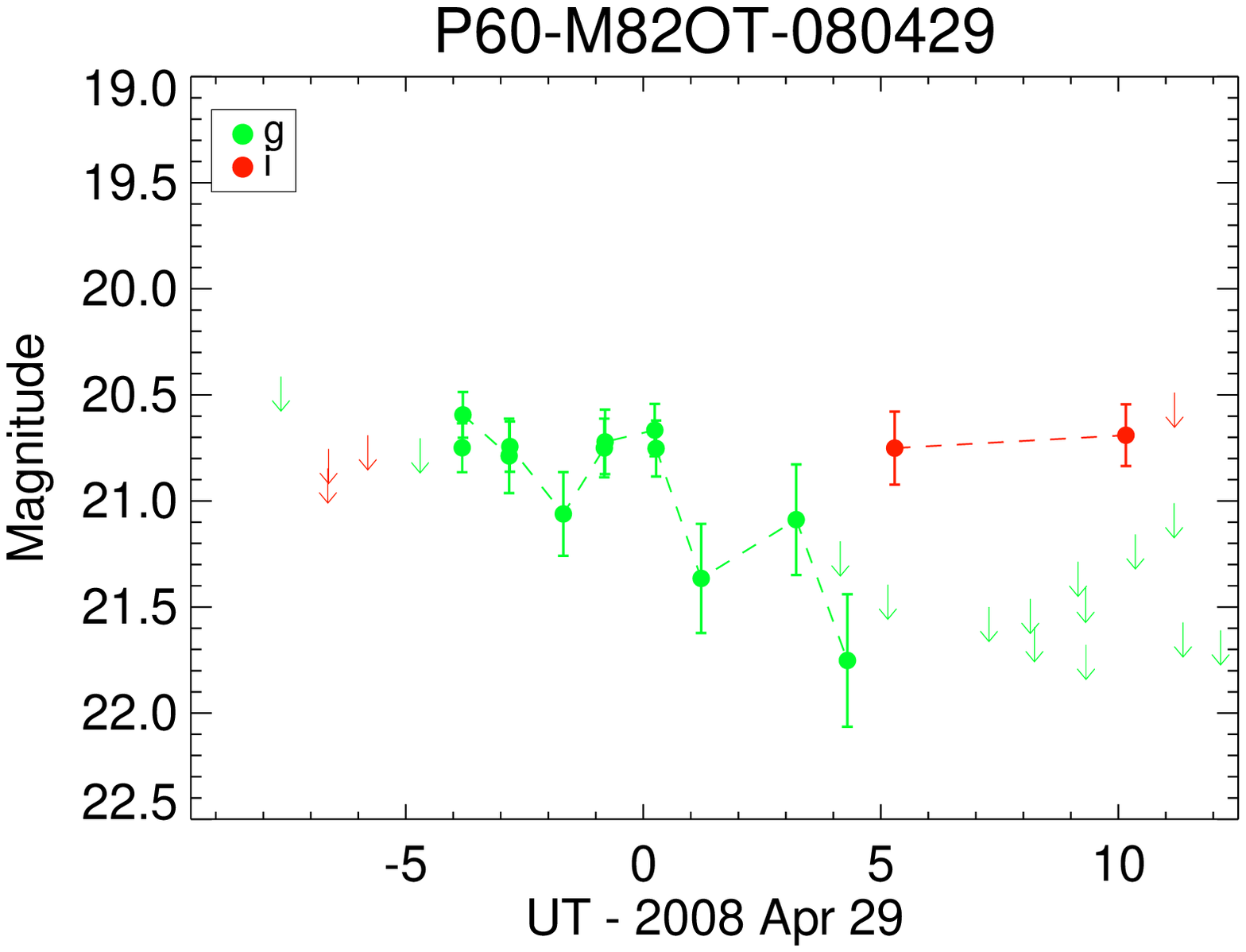,width=0.48\linewidth, angle=0} \\
\end{tabular}
\caption{\small Lightcurves of novae in M82 and NGC2403. Note that P60-M82OT-081119 
is much redder than typical novae.
\label{fig:m82lc}}
\end{center}
\end{figure*}

\begin{figure*}[!hbt]
\begin{center}
\begin{tabular}{cc}
\psfig{figure=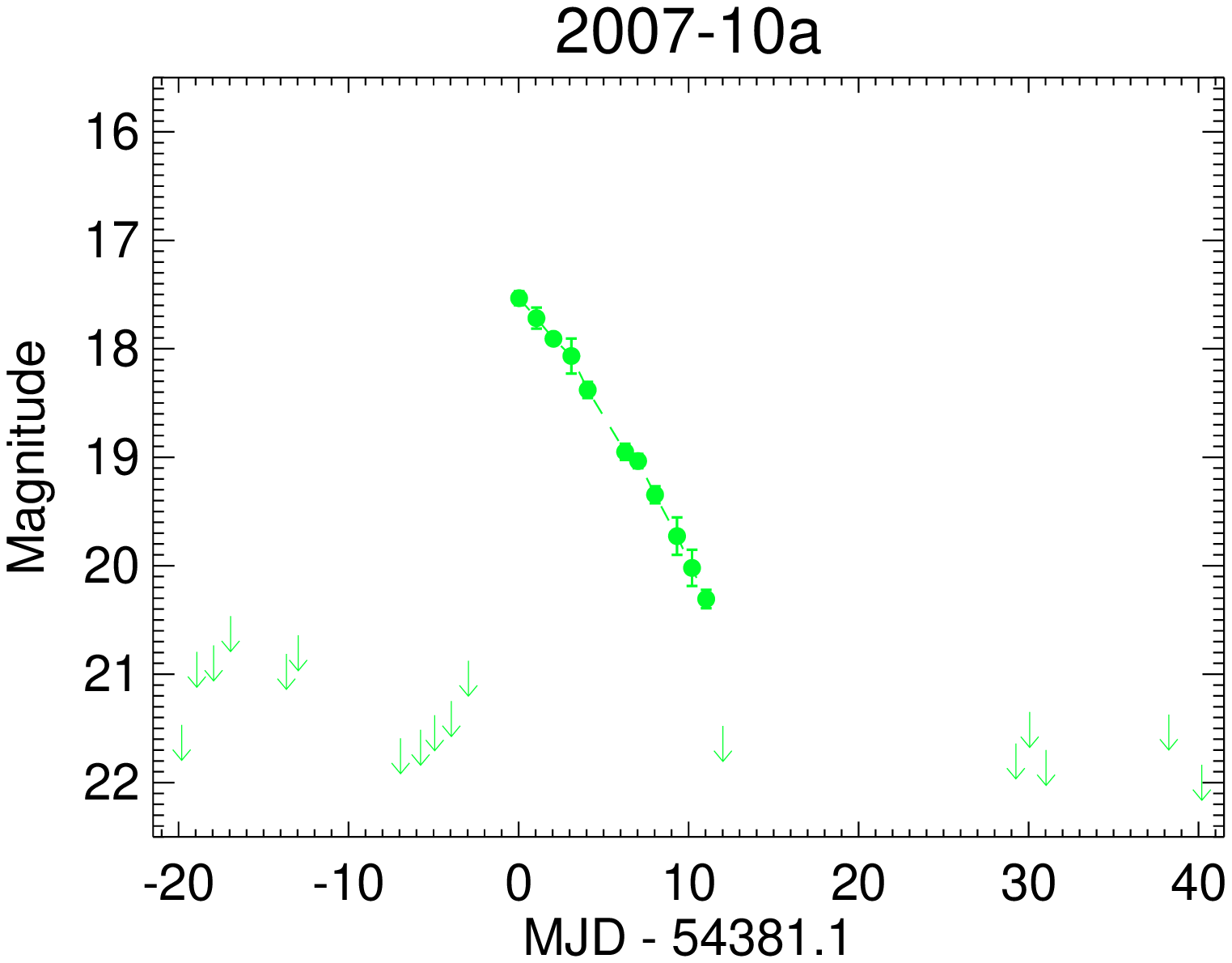,width=0.48\linewidth, angle=0} &
\psfig{figure=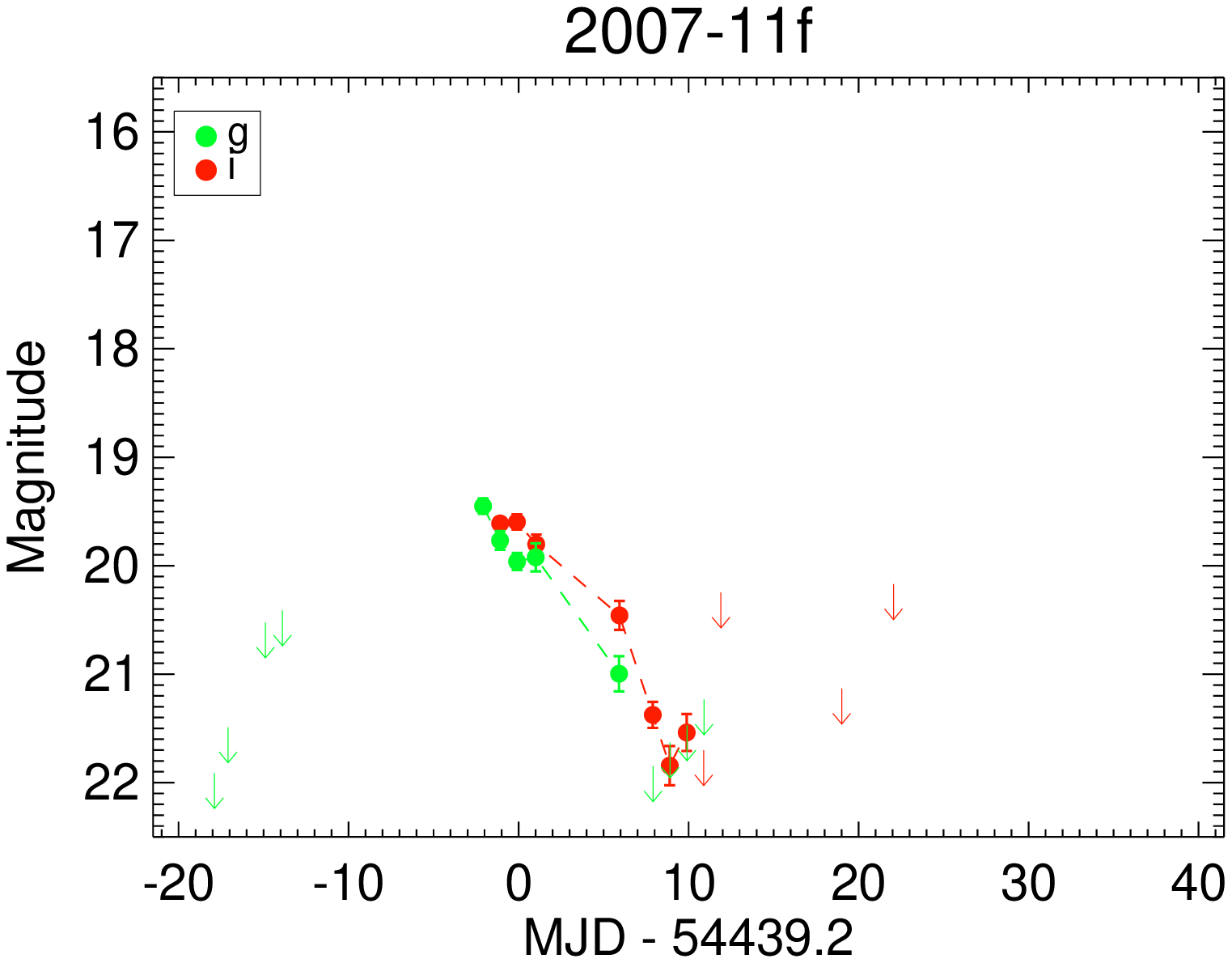,width=0.48\linewidth, angle=0} \\
\psfig{figure=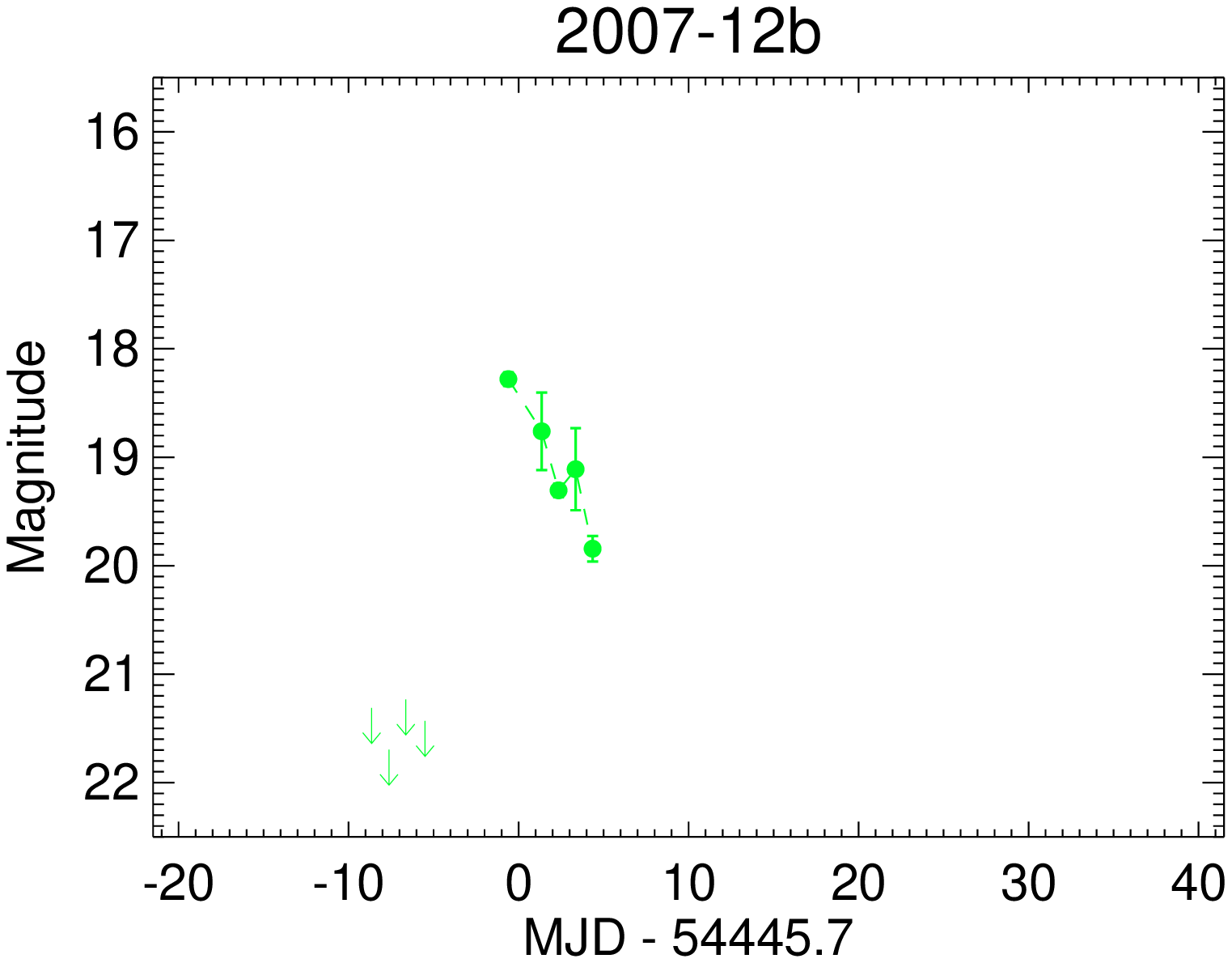,width=0.48\linewidth, angle=0} &
\psfig{figure=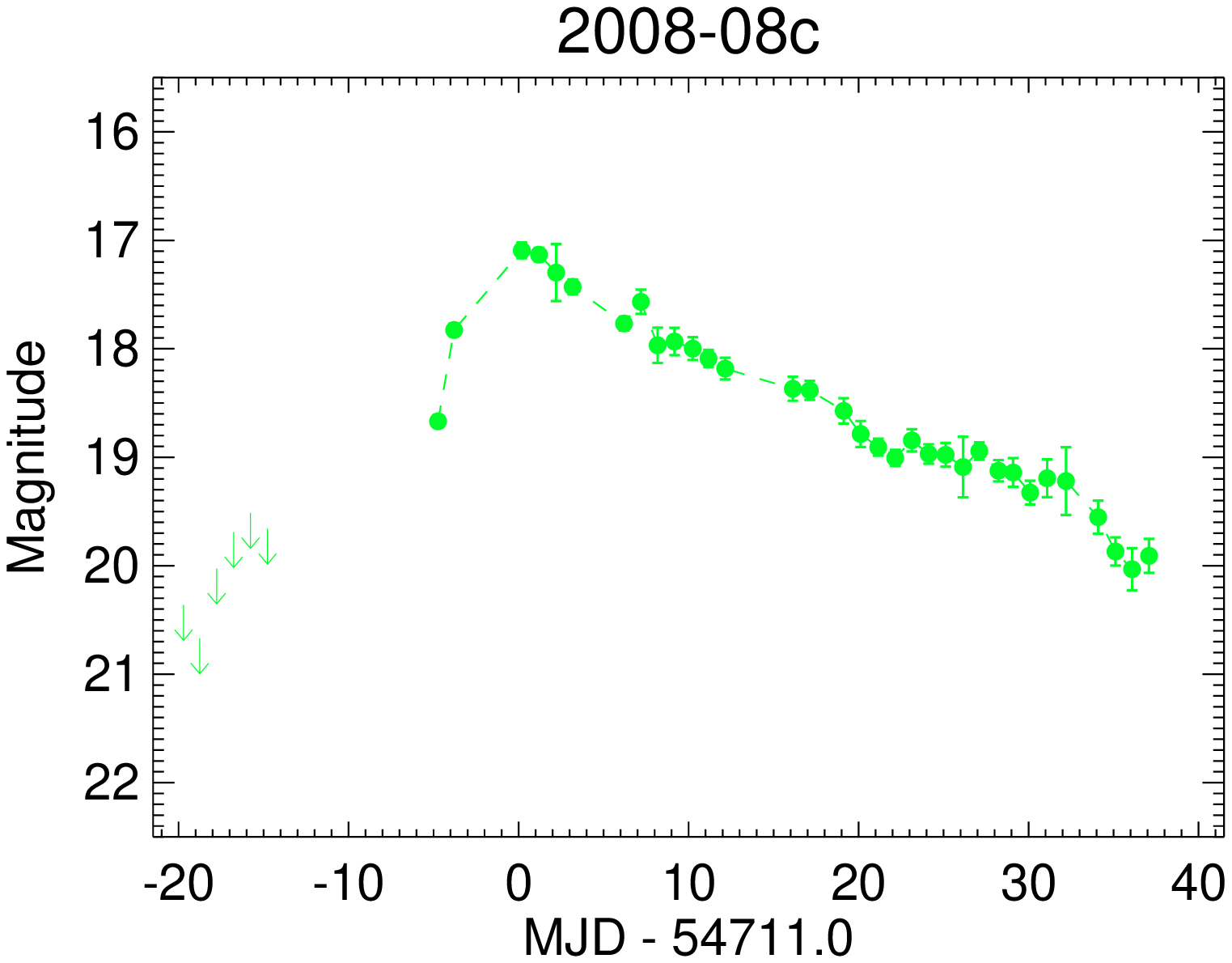,width=0.48\linewidth, angle=0} \\
\psfig{figure=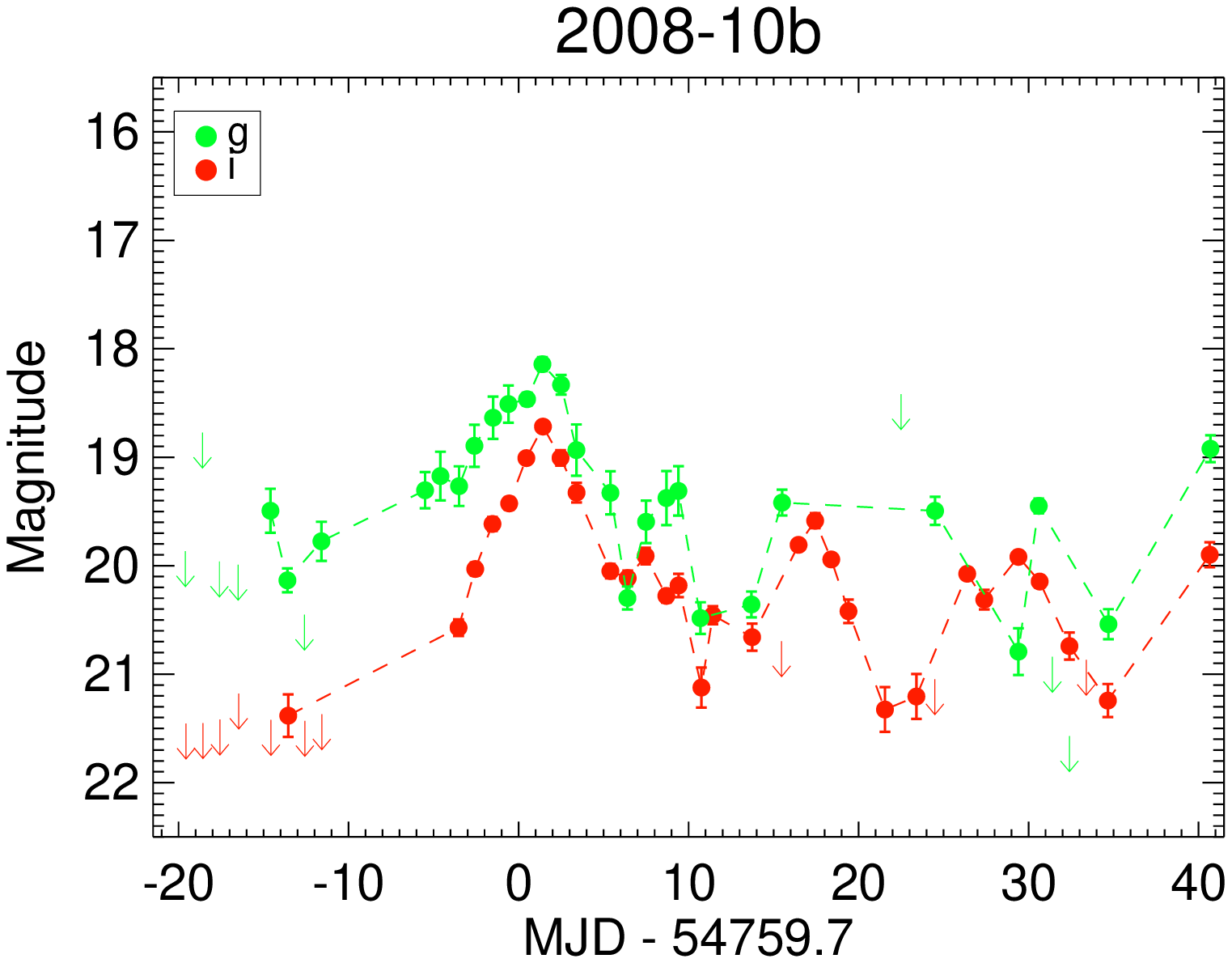,width=0.48\linewidth, angle=0} &
\psfig{figure=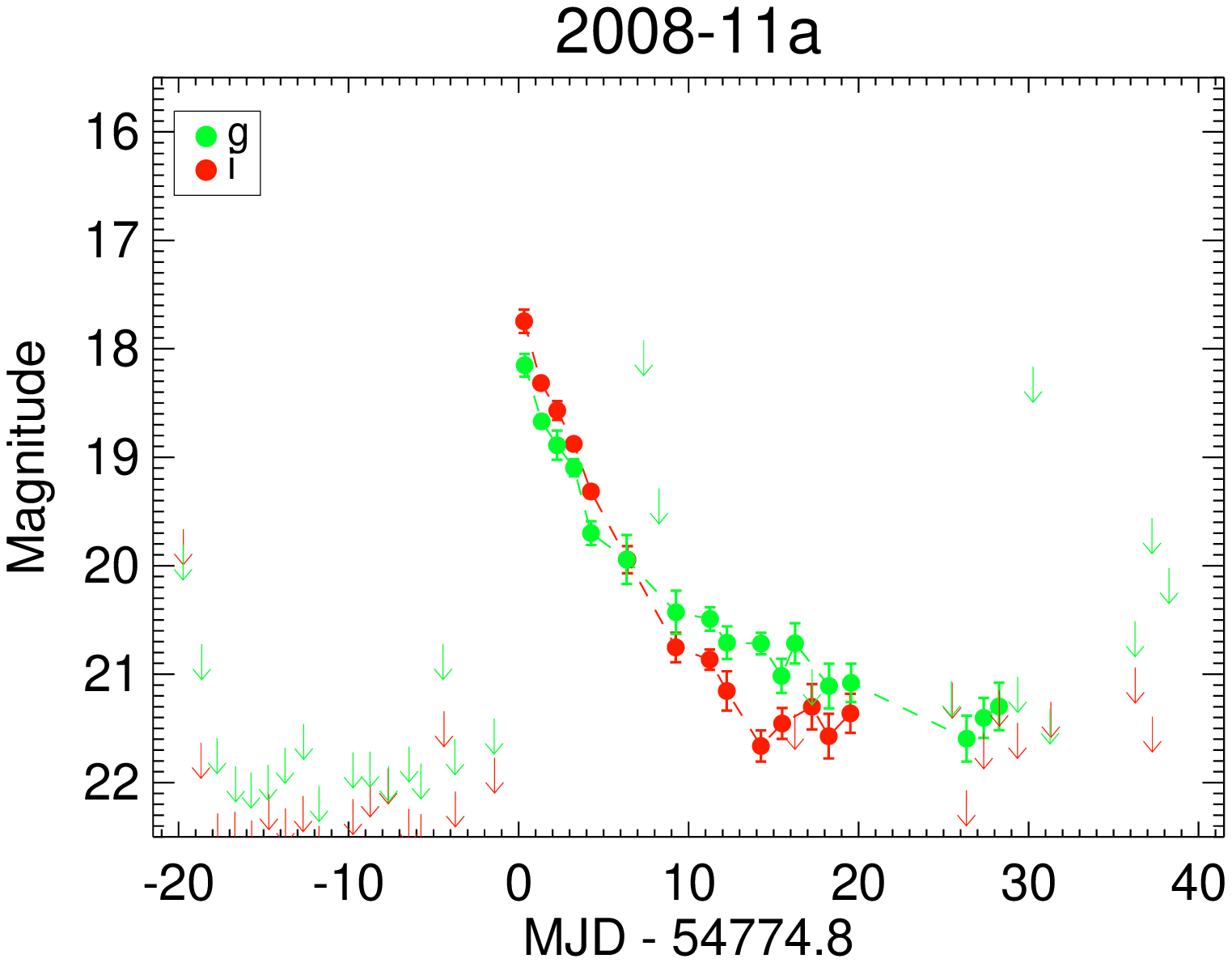,width=0.48\linewidth, angle=0} \\
\end{tabular}
\caption{\small Lightcurves of additional novae in M31.
\label{fig:morem31}}
\end{center}
\end{figure*}

\begin{figure}[!hbt]
\begin{center}
\psfig{figure=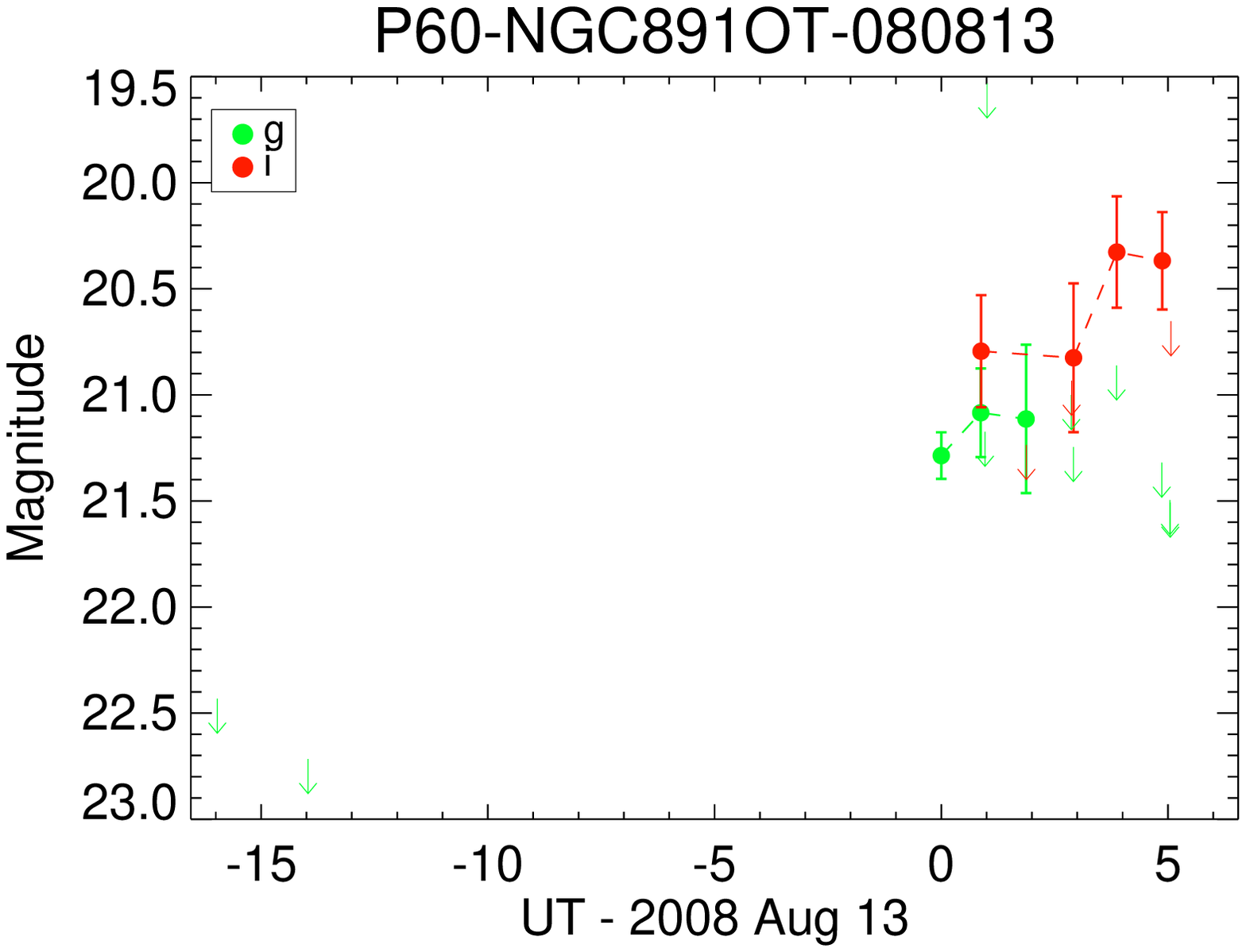,width=0.9\linewidth, angle=0}
\caption{\small Lightcurve of possible nova in NGC\,891. This was not spectroscopically confirmed 
and is not used in the MMRD analysis. 
\label{fig:mosaic}}
\end{center}
\end{figure}

\subsection{Spectroscopy}
An integral part of P60-FasTING was follow-up spectroscopy to confirm 
and classify discovered candidate transients. Since we were looking for
fast evolving phenomenon, we triggered our Target Of Opportunity program on
the Keck I and Palomar Hale telescopes soon after discovery. Sometimes due to bad 
weather or bright moon-phase\footnote{low resolution spectrographs are usually available 
on both these telecopes only during the dark fortnight}, neither of these was an option. We 
resorted to the queue-scheduled service-observed programs on Gemini or HET telescopes. 
A log of spectroscopic observations can be found in Table~\ref{tab:spec}.

We emphasize that spectroscopy is crucial in distinguishing between an optical transient
which happened to be co-incident with a nearby galaxy and a classical nova. 
For instance, we took spectra of several optical transient candidates which did not turn out to be novae:  
a foreground M-dwarf flare in the Milky Way spatially coincident with NGC\,7640;  
a background supernova; a luminous blue variable in NGC\,925.

\begin{figure*}[!hbt]
\begin{center}
\psfig{figure=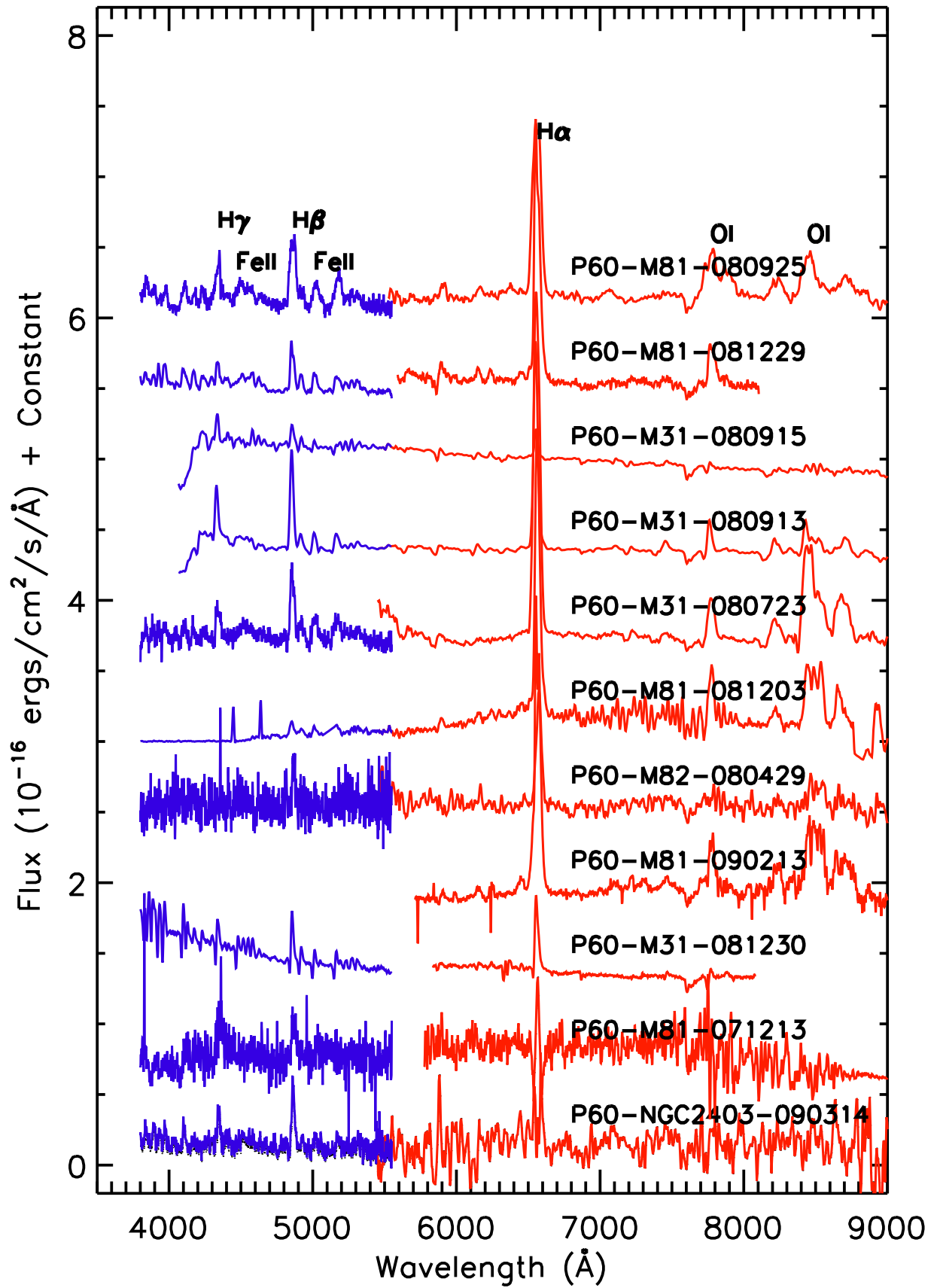, angle=0}
\caption{\small Optical spectra of P60-FasTING novae. Majority are Fe II class. Note that they have
been arbitrarily offset along the y-axis for clarity. 
\label{fig:spec}}
\end{center}
\end{figure*}

Data were reduced in \texttt{iraf} using standard tasks in the NOAO package \texttt{onedspec}
and the spectra are shown in Figure~\ref{fig:spec}.


\section{Analysis}
\label{sec:analysis}
The primary photometric analysis was to measure the peak absolute magnitude and rate of decline.
The peak magnitude had to be corrected for extinction using spectra or colors. The primary
spectroscopic analysis was to classify the spectra. 

\subsection{Extinction}
A multitude of methods have been used in the literature to measure
extinction to novae. \citet{dbk+06} compared a synthetic dust-free stellar
$r-i$ map of M31 to an observed $r-i$ color map of M31 and used the difference
between the maps to generate a dust map of M31. The location of the nova on this map determined
how much extinction needed to be applied. This assumed that the novae were behind the
galaxy and suffered extinction due to the entire column of dust. The average extinction 
as determined by this method is A$_{i}$=0.8. The galactic extinction along the
line of sight of M31 of A$_{i}$=0.13. 
\citet{srq+09} compared the observed color of the new nova to that of a well-studied nova to
derive the extinction. \citet{K61} used the Balmer decrement ($H_{\alpha}/H_{\beta}$)
and attributed the excess in the ratio over the theoretical Case B value to dust. 

Our preferred method of computing extinction is by using the spectroscopic
Balmer decrement where nebular spectra are available. We subtract the continuum,
measure the flux ratio of the two lines, and then use:
$$A_{g} = 3.793 \times E(B-V) = 3.650 \times E(g-r) $$
$$   = 3.650 \times \log_{10}(H_{\alpha}/H_{\beta}) - 1.75 \pm 0.14$$
The uncertainty comes from the range in expected ratios for Case B of 2.76--3.30. 

Our second choice is to use the $g-i$ color of the nova at maximum, compare against 
the typical $g-i$ color and attribute the reddening to dust. 

\citet{vy87} compiled photometry of several Galactic novae and derived an average color
at maximum of $\langle B-V \rangle_{0}$=0.23 $\pm$ 0.06\,mag. Following \citet{srq+09}, we 
can use the colors of an A5V star (T=8200\,K) to translate $\langle B-V \rangle_{0}$ 
to $\langle g-i \rangle_{0}$. Using colors of an A5V star from \cite{kh07}, we get:
 $$\langle g-i \rangle_{0} = 1.88-2.15 = -0.27$$\,mag. Now, $$A_{g} = 3.793 \times E(B-V) = 2.223 \times E(g-i) $$.
Hence, $$A_{g} = 2.223 \times [(g-i)_{obs}- \langle g-i \rangle_{0}] =  2.223 \times [(g-i)_{obs} + 0.27]$$

For the case of M31N2007-11d \citep{srq+09}, where data for both the above options 
are available, we get consistent answers: $g-i$=0.7\,mag at maximum suggests A$_{g}$=0.75\,mag and 
a Balmer decrement of 4.6 suggests A$_{g}$=0.65\,mag. 


If neither a color at maximum nor a spectrum at late-time is available, 
we use the average line-of-sight extinction to the host galaxy using 
the Schlegel maps. The uncertainty in extinction calculation significantly
contributes to the uncertainty in the peak magnitude. 

We note here that for the case of P60-M82OT-081119, the light curve was unusually
red for a nova and the extinction correction may be overestimated. 



\subsection{Rate of decline}
The heterogenity in nova light curves suggests that a single
parameter may not characterize the decline well. Traditionally, the time
to decay from peak by one magnidue (t1), two magnitudes (t2) or three magnitudes (t3) is 
used.  For several novae (e.g. M31N2007-10a, M31N2008-08c, M31N2008-11a), the decline is more or less linear 
and t1 can be approximated as half of t2. For some novae (e.g. M31N2008-10b), the light curve 
behavior is more complex and this simplification is not applicable. In Table~\ref{tab:mmrd}, we see that 
t2 values (where available) are sometimes larger and sometimes smaller than twice the t1 value.

We note here that we did not have data to measure the decline of the nova in NGC\,891 and hence it is
excluded from further MMRD analysis.

\subsection{Rate of rise}
Given the cadence of P60-FasTING, we were able to catch several novae on the rise. We define
the rate of rise as the average slope between first detection and peak detection and summarize
in Table~\ref{tab:mmrd}. We find a wide range of rise-times, from $>$1.8\,mag\,day$^{-1}$ (e.g. M31N2008-11a) to 
0.2\,mag\,day$^{-1}$ (e.g. M31N2008-09a). It is not clear how previous determinations of the MMRD in the literature 
dealt with the uncertainty in the peak magnitude due to inadequate coverage. Especially since previous surveys 
likely had a relatively slower cadence, missing the peak may be a substantial source of error.
Slower cadence and/or shallower depth would correspond to a weaker constraint on the rise time of the nova.
Due to gaps on account of weather, some of the P60 light curves have constraint weaker than 0.1\,mag~day$^{-1}$ 
on the rate of rise. Hence, we do not use the lightcurves of P60-M81OT-080926 or P60-M82OT-080429 for subsequent 
analysis of the MMRD relation.

\begin{table*}[!hbt]
\begin{center}
\caption[]{\bf Spectroscopy}
\begin{tabular}{llllllll}
\hline
\hline
Transient Name & Spectroscopy Date & Telescope & Instrument & Classification & Observer \cr
\hline
P60-NGC2403-090314 & 2009 Mar 20.145 & P200 & DBSP \citep{og82} & Fe Class? & Kasliwal,Ellis \cr
P60-M81-090213 & 2009 Feb 18.510 & Keck I & LRIS \citep{occ+95} & Fe Class & Ofek \cr
P60-M81-081229 & 2008 Dec 31.40  & P200   & DBSP \citep{og82} & Fe Class & Rau,Salvato \cr
P60-M31-081230 & 2008 Dec 31.104 & P200   & DBSP \citep{og82} & Fe Class & Rau,Salvato \cr
P60-M81-081203 & 2008 Dec,4,5,16 & P200,Gemini  & DBSP,GMOS-N \citep{hja+04}  & Fe Class & Kasliwal \cr
P60-M81-080925 & 2008 Sep 29.51 & P200 & DBSP \citep{og82} & Fe Class & Quimby \cr
P60-M31-080915 & 2008 Sep 20.2 & HET & LRS \citep{hnm+98} & Fe Class & Shafter \cr
P60-M31-080913 & 2008 Sep 22.4 & HET & LRS \citep{hnm+98} & Fe Class & Shafter \cr
P60-M31-080723 & 2008 Aug 1 & P200 & DBSP \citep{og82} & Fe Class & Ofek \cr
P60-M82-080429 & 2008 May 2.28 & P200 & DBSP \citep{og82} & \nodata & Cenko \cr
P60-M81-071213 & 2007 Dec 15.565 & Keck & LRIS \citep{occ+95} &\nodata & Ofek \cr
\hline
\hline
\label{tab:spec}
\end{tabular}
\end{center}
\end{table*}

\begin{table*}[!hbt]
\begin{center}
\caption[]{\bf Nova Characteristics}
\begin{tabular}{lllllllll}
\hline
\hline
Nova Name & Balmer Decrement & Spectral Phase & Color at Peak  & Extinction & Rate of Rise & Abs-Mag & t1 & t2 \cr
          & F{H${\alpha}/$H${\beta}$} & & $g-i$(mag) & A$_{g}$(mag) & mag~day$^{-1}$ & M$_{g}$(mag) & days & days \cr
\hline
P60-NGC2403-090314 & 5.0     & Nebular  & \nodata &  0.8    & 1.3    & $-$9.0    & 3.3  & $>$6    \cr 
P60-M82-090314     & \nodata & \nodata  & $-$0.3  &  0.6    & $>$1.2 & $-$8.5    & 2.4? & $>$3.3  \cr
P60-M81-090213     &  \nodata       & Nebular  &  0.7    &  2.2    & 0.2    & $-$9.9    & 5?   & $>$10.9 \cr
P60-M31-081230 (2008-12b) &  \nodata       & \nodata        & $-$0.17 &  0.24   & 0.6    & $-$7.5    & 12.3 &  \cr
P60-M81-081229     & 2.4     & Near-Max &  0.1    &  0.90   & $>$0.1 & $-$8.7    & 2.9  & \nodata \cr 
P60-M81-081203     &  \nodata       &  \nodata       & $-$0.03 &  0.53   & 0.7    & $-$8.0    & 7.5? & $>$23   \cr
P60-M82-081119     & \nodata & \nodata  &  0.98   &  2.8    & 0.6    & $-$10.7   & 4.0  & \nodata \cr 
P60-M81-081027     & \nodata & \nodata  & \nodata &  0.3    & 0.2    & $-$7.6    & 4.0  & \nodata \cr 
P60-M81-080926     & 1.4     & Near-Max & -0.2    &  0.13   &\nodata & $-$8.5    & 8.9  & 14.0    \cr 
P60-M31-080915 (2008-09c)    & 1.4     & Near-Max & $-$0.72 &  0.24   & 0.4    & $-$7.8    & 9.1  & 16.6    \cr
P60-M31-080913 (2008-09a)    & 2.5     & Near-Max & $-$0.30 &  0.24   & 0.2    & $-$6.8    & 6.3  & 16.0    \cr
P60-M31-080723 (2008-07b)   & 14.3    & Nebular? & $-$0.20 &  2.5    & 0.2    & $-$7.6    & 5.0  & 12.0    \cr 
P60-M82-080429     & 6.3     & Nebular? & \nodata &  1.2    &\nodata & $-$8.5    & 8.1  & \nodata \cr 
P60-M81OT-071213   & 3.8     & Nebular? &  0.5    &  0.4    & $>$0.6 & $-$7.8    & 1.0  & \nodata \cr 
2007-10a           & \nodata & \nodata  & \nodata & $>$0.24 & $>$1.2 & $-$7.0    & 4.1? & 8.6     \cr 
2007-11f           & \nodata & \nodata  & $-$0.16 & 0.24    & $>$0.1 & $-$5.1    & 5.0  & $>$8.0  \cr 
2007-12b           & \nodata & \nodata  & \nodata & $>$0.24 & $>$0.6 &  $-$6.3   & 3.5  & $>$5.0  \cr
2008-08c           & \nodata & \nodata  & \nodata & $>$0.24 & 0.3    & $-$7.5    & 11.0 & 26.3    \cr
2008-10b           & \nodata & \nodata  & $-$0.58 & $>$0.24 & 0.2    & $-$6.5    & 6.0  & 12.3?   \cr
2008-11a           & \nodata & \nodata  &  0.41   &  1.5    & $>$1.8 & $-$7.7    & 2.9  & 7.5     \cr
\hline             
\hline            
\label{tab:mmrd}   
\end{tabular}      
\end{center}       
\end{table*}
\bigskip

\begin{figure}[!hbt]
\begin{center}
\epsfig{figure=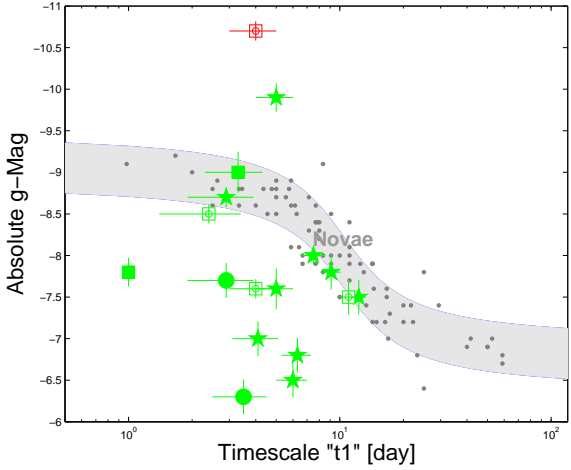,width=\linewidth, angle=0}
\caption{\small Maximum magnitude (absolute $g$-band) versus rate of decline (time to decay from peak by 1 mag). Gray region denotes the \citet{dl95} MMRD relation and dark gray dots the nova sample they used. The P60-FasTING sample is shown with symbols that denote spectral type --- Fe II class (star), He/N class (circle), spectrum with no prominent features for classification (squares) and no spectrum (empty square).    
\label{fig:taumv}}
\end{center}
\end{figure}
\begin{figure}[!hbt]
\begin{center}

\epsfig{figure=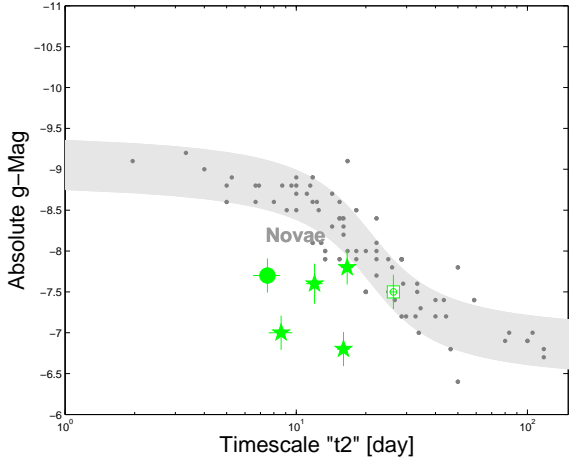,width=\linewidth, angle=0}
\caption{\small Maximum magnitude (absolute $g-$band) versus rate of decline (time to decay from peak by 2 mag). Only the six novae with the best sampled light curves are shown.  
\label{fig:taumv_t2}}
\end{center}
\end{figure}

\begin{figure}[!hbt]
\begin{center}
\epsfig{figure=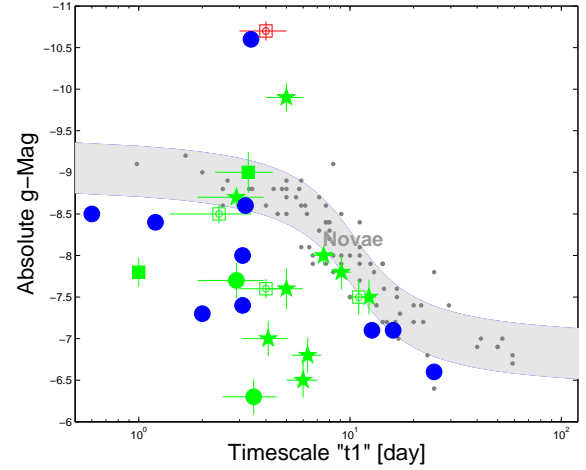,width=\linewidth, angle=0}
\caption{\small Comparison of the P60-FasTING nova sample (green symbols) with the Galactic recurrent novae (blue circles, data from \citet{s09}). 
\label{fig:taumvRN}}
\end{center}
\end{figure}

\begin{figure}[!hbt]
\begin{center}
\epsfig{figure=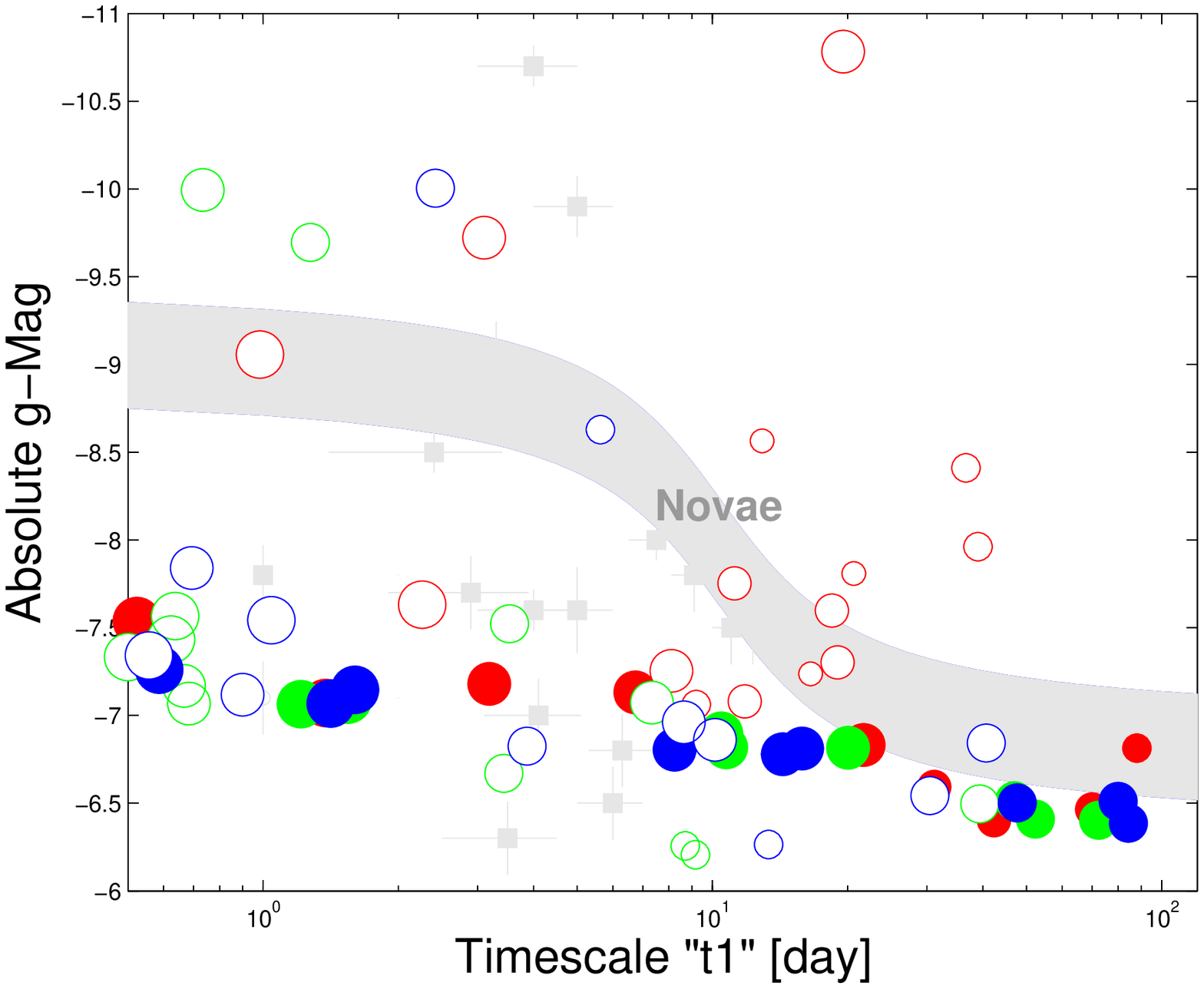,width=\linewidth, angle=0}
\caption{\small Theoretical results of \citet{yps+05} over a wide range of nova parameters --- we use the colors of an A5V star to convert L4max to M$_{g}$ and approximate t1 as t$_{ml}$/3.  The size of the symbol is proportional to the mass of the white dwarf. The color of the symbol denotes temperature --- 10 million K (red), 30 million K (green), 50 million K (blue). Empty circles denote lower accretion rate in the range 10$^{-12.3}$---10$^{-10}$M$_{\odot}$yr$^{-1}$ and filled circles denote higher accretion rate in the range 10$^{-9}$---10$^{-6}$M$_{\odot}$yr$^{-1}$. Note that the density of circles is unrelated to 
the relative populations. 
\label{fig:taumv_theory}}
\end{center}
\end{figure}

\subsection{Spectral Classification}
For spectroscopy the primary analysis was to classify the novae
by their spectra. The taxonomy of novae were laid out by  \cite{p64}
and \cite{m60}. The most prominent feature in all classical novae is Balmer emission.
\citet{w92} propose that there is a two component structure of the emitting gas --
discrete shell and continuous wind. If the wind mass-loss rate is low, the effective photosphere
is smaller, the radiation temperature higher, the level of ionization of the shell higher, 
resulting in a shell-dominated He/N spectrum. If the wind mass-loss rate is high, it results 
in a wind-dominated Fe II spectrum.  

Thus, classical novae are divided into two principal families ---  
the ``Fe'' class (dominated by Fe~II lines, often low velocity and showing P-Cygni profiles) 
and ``He/N'' class (dominated by He and N lines,
often high velocity and flat or jagged-topped profiles). These evolve into nebular spectra
with four classes based on the prominent forbidden lines --- standard (e.g. [N II], [O II], [O III]), 
neon (e.g. [Ne V], [Ne III]), coronal (e.g.  [Fe X]) or no forbidden lines. The Fe class
novae are expected to evolve into standard or neon nebular spectra. The He/N class
are expected to evolve into neon, coronal or no forbidden line spectra.
Some novae are classified as ``hybrid'' as they start out with high velocity Fe II features
and quickly evolve into showing He/N features (e.g. V745 Sco, V3890 Sgr, M31N2008-11a). 

Majority of the P60-FasTING spectra show clear permitted lines from Fe II (42), Fe II (37,38)
and O I. The line velocities are low and typical Gaussian FWHM are $<$2500\,km s$^{-1}$ with the
exception of P60-M81-080925 where H$_{\alpha}$ velocity is 3000\,km s$^{-1}$. P60-NGC2403-090314 shows
weak Fe II(42) and weak P-Cygni profiles in Balmer lines and can tentatively also be classified
as Fe II class. P60-M81-071213 and P60-M82-080429 have very low SNR and no feature other than 
the Balmer lines are detected, hence, we cannot classify them. Multiple spectra of 
P60-M81-081203 were taken --  initially, the
spectra show a featureless continuum (obtained a few days prior to maximum light) and 
later (about a week after maximum light), evolved to show Balmer lines, Fe II (42), O I. We 
summarize spectral classifications in Table~\ref{tab:spec}.
For four novae in M31, other groups obtained spectra and we summarize their classification
in Table~\ref{tab:addm31novae}.


\section{Discussion}
\label{sec:discussion}
In comparison to traditional nova searches, P60-FasTING was designed as 
a faster cadence and deeper survey. Weather-permitting, galaxies in the sample
were imaged every night to a mean depth of Gunn-$g\,<\,21$\,mag. 
Hence, P60-FasTING was sensitive to transients that are less luminous and evolve faster
than classical novae. 

Given that our light curves are well-sampled, and we have spectra or
color measurements to correct for extinction, we can securely measure both
the maximum magnitude and the rate of decline. To our surprise, as demonstrated
in Figure~\ref{fig:taumv}, we find that the P60-FasTING nova sample is evidently 
inconsistent with the MMRD relation \citep{dl95}. 

In Figure~\ref{fig:taumv}, the 
decay time is measured as the time to decay by one magnitude. To test whether 
the apparent photometric diversity is consistent with the MMRD over a longer timescale, we
plot the time to decay by two magnitudes in Figure~\ref{fig:taumv_t2}.   
Furthermore, we restrict this to the sub-sample of six classical novae in M31 
with the best-sampled light curves (see lightcurves of P60-M31OT-080915, 
P60-M31OT-080913, P60-M31OT-080723, M31N2007-10a, M31N2008-08c,M31N2008-11a
in Figure~\ref{fig:m31lc} and Figure~\ref{fig:morem31}). Even this sub-sample does not obey the
MMRD relation. This scatter is larger than the $\pm$0.8\,mag predicted on theoretical grounds 
by \citet{s81}.

Despite the atypical photometric signature, the P60-FasTING nova sample shows no 
spectroscopic peculiarities. In Figure~\ref{fig:taumv},
the symbols indicate the spectral class --- majority are Fe II class (stars), a couple 
are He/N class (circles), some have spectra with no prominent features for classification 
(filled squares) and a few have no spectra (empty squares). 

We could hypothesize that some of the P60-FasTING novae are not classical but recurrent (classical novae 
which recur on a timescales shorter than a century) since recurrent novae are also
known not to obey the MMRD relation. Recurrent novae are expected to occur in the most 
massive white dwarfs with high accretion rates. A small amount of mass accreted on a short 
timescale is sufficient to trigger thermonuclear runaway. Recently, \citet{s09} compiled all available photometry 
over the past century on the ten recurrent novae in our galaxy --- overplotted to compare with
the P60-FasTING sample in Figure~\ref{fig:taumvRN}. 

There are three ways to test the recurrent nova hypothesis. First, spectra of recurrent novae 
have high velocities and belong to either He/N or hybrid class 
(e.g. V3890 Sgr, V745 Sco, V394 CrA in \citealt{w92} and \citealt{wph94}).
We find that majority off the P60-FasTING novae do not share the spectroscopic properties
of recurrent novae. P60-FasTING novae mostly belong to the Fe II class and have low velocities. Second, recurrent
novae are often a few magnitudes brighter than classical novae at quiescence. \citet{s09} suggest
that recurrent novae range from $-$4.1 $<$ M$_{V} <$ 3.2 and classical novae range from 1.1 $<$ M$_{V}<$ 7.0
at quiescence. Given the distance modulus to these galaxies, this test is within reach of 10-m class telescopes
and easy with HST (e.g. see \citealt{bds+09}). Third, the unambiguous test of whether an eruption is recurrent is to continue to monitor these 
galaxies for the next few decades until another eruption is witnessed.  

In order to decipher the nature of this new sub-class of novae we
turn to the fundamental physics of classical novae. The physics is governed by four parameters 
- mass of the white dwarf, temperature, accretion rate and composition. The MMRD relation is explained
with the mass of the white dwarf being the single, dominant parameter. Perhaps, the P60-FasTING sample 
of faint and fast novae can be explained based on an unexplored region of this four-parameter 
phase space?
Could some P60-FasTING novae come from hot and massive white dwarfs? 
If it is hot, then the thermonuclear runaway would not be as explosive and thus, the peak luminosity 
would be fainter. Also, the higher temperature would result in a smaller amount of
envelope mass being sufficient to trigger thermonuclear runaway and thus, the timescale 
would be faster. 

Recent theoretical efforts have explored nova diversity (e.g. \citealt{tb04}, \citealt{sb09}, \citealt{eyk+07}).
\citet{yps+05} present an extended grid of nova models to explore a wider parameter space
(in mass, temperature and accretion rate) than traditionally explored for classical novae 
subject to physical constraints (such as conditions for thermonuclear runaway). 
We summarize the results of 
the variety of models they run in Figure~\ref{fig:taumv_theory}. Some hot and massive white 
dwarfs with high accretion rates can result in a faint and fast nova population consistent 
with the P60-FasTING sample. Indeed, \citet{yps+05} predict the existence of remarkably 
small amplitude novae across the entire span of decay rates.

Finally, we note that more than half of the P60-FasTING nova sample is inconsistent
with the MMRD. This suggests that faint and fast novae are commonplace and cannot
be explained by a rare type of white dwarf.

\section{Conclusion}
\label{sec:conclusion}
We conclude that P60-FasTING has uncovered classical novae in a new region in the 
luminosity-timescale phase space of optical transients. Classical novae span at least two orders
of magnitude in time and two orders of magnitude in luminosity. Future surveys would have 
a large enough sample to meaningfully constrain the relative populations of classical novae in 
the different areas of phase space.

P60-FasTING was designed as a pilot project, to begin to set the stage for future projects such as 
Palomar Transient Factory (PTF\footnote{http://www.astro.caltech.edu/ptf}, \citealt{lkd+09}, \citealt{rkl+09}, \citealt{gsv+08}), PanSTARRS (PS1\footnote{http://pan-starrs.ifa.hawaii.edu}) and Large Synoptic Survey Telescope (LSST\footnote{http://www.lsst.org}). 
Both PTF and PS1 are now underway. PTF is looking at several nearby galaxies with a similar 
depth and cadence as P60-FasTING. Among nearby galaxies, PS1 day-cadence fields only cover 
M31 but are a couple of magnitudes deeper. LSST will be both deeper and faster cadence and cover 
the visible sky.  P60-FasTING is only the trailblazer for the uncovering of a wealth of information about 
classical novae by near-future synoptic surveys.

\bigskip 
\begin{acknowledgements}
We thank Marina Orio for discussion, in particular, the suggestion that some
of the faint novae may be recurrent novae. We thank A. Shafter, M. Shara, L. Bildsten
and O. Yaron for valuable feedback. We are grateful to A. Becker for
making his software \texttt{hotpants} and \texttt{wcsremap} available for public 
use. MMK thanks the Gordon and Betty Moore Foundation for the Hale Fellowship in 
support of graduate study. S.B.C. is grateful for generous support from Gary and 
Cynthia Bengier, the Richard and Rhoda Goldman Fund, and National Science Foundation (NSF) grant AST0908886.
MMK thanks the Palomar Observatory staff for their help
in maximizing the efficiency and image quality of the Palomar 60-inch. Some of the data 
presented herein were obtained at the W.M. Keck Observatory, which is operated as a 
scientific partnership among the California Institute of Technology, the University 
of California and the National Aeronautics and Space Administration. The Observatory was made 
possible by the generous financial support of the W.M. Keck Foundation.
The authors wish to recognize and acknowledge the very significant 
cultural role and reverence that the summit of Mauna Kea has always had 
within the indigenous Hawaiian community.  We are most fortunate to have 
the opportunity to conduct observations from this mountain.

\end{acknowledgements}





\begin{thebibliography}{55}
\expandafter\ifx\csname natexlab\endcsname\relax\def\natexlab#1{#1}\fi

\bibitem[{{Arp}(1956)}]{a56}
{Arp}, H.~C. 1956, \aj, 61, 15

\bibitem[{{Barsukova} {et~al.}(2008){Barsukova}, {Fabrika}, {Hornoch},
  {Sholukhova}, \& {Valeev}}]{bfh+08}
{Barsukova}, E., {Fabrika}, S., {Hornoch}, K., {Sholukhova}, O., \& {Valeev},
  A. 2008, The Astronomer's Telegram, 1871, 1

\bibitem[{{Bode} {et~al.}(2009){Bode}, {Darnley}, {Shafter}, {Page},
  {Smirnova}, {Anupama}, \& {Hilton}}]{bds+09}
{Bode}, M.~F., {Darnley}, M.~J., {Shafter}, A.~W., {Page}, K.~L., {Smirnova},
  O., {Anupama}, G.~C., \& {Hilton}, T. 2009, \apj, 705, 1056

\bibitem[{{Darnley} {et~al.}(2006){Darnley}, {Bode}, {Kerins}, {Newsam}, {An},
  {Baillon}, {Belokurov}, {Calchi Novati}, {Carr}, {Cr{\'e}z{\'e}}, {Evans},
  {Giraud-H{\'e}raud}, {Gould}, {Hewett}, {Jetzer}, {Kaplan},
  {Paulin-Henriksson}, {Smartt}, {Tsapras}, \& {Weston}}]{dbk+06}
{Darnley}, M.~J., {Bode}, M.~F., {Kerins}, E., {Newsam}, A.~M., {An}, J.,
  {Baillon}, P., {Belokurov}, V., {Calchi Novati}, S., {Carr}, B.~J.,
  {Cr{\'e}z{\'e}}, M., {Evans}, N.~W., {Giraud-H{\'e}raud}, Y., {Gould}, A.,
  {Hewett}, P., {Jetzer}, P., {Kaplan}, J., {Paulin-Henriksson}, S., {Smartt},
  S.~J., {Tsapras}, Y., \& {Weston}, M. 2006, \mnras, 369, 257

\bibitem[{{della Valle} \& {Livio}(1995)}]{dl95}
{della Valle}, M. \& {Livio}, M. 1995, \apj, 452, 704

\bibitem[{{Di Mille} {et~al.}(2008){Di Mille}, {Ciroi}, {Orio}, {Rafanelli},
  {Bianchini}, {Nelson}, \& {Andreuzzi}}]{dco+08}
{Di Mille}, F., {Ciroi}, S., {Orio}, M., {Rafanelli}, P., {Bianchini}, A.,
  {Nelson}, T., \& {Andreuzzi}, G. 2008, The Astronomer's Telegram, 1818, 1

\bibitem[{{Downes} \& {Duerbeck}(2000)}]{dd00}
{Downes}, R.~A. \& {Duerbeck}, H.~W. 2000, \aj, 120, 2007

\bibitem[{{Epelstain} {et~al.}(2007){Epelstain}, {Yaron}, {Kovetz}, \&
  {Prialnik}}]{eyk+07}
{Epelstain}, N., {Yaron}, O., {Kovetz}, A., \& {Prialnik}, D. 2007, \mnras,
  374, 1449

\bibitem[{{Ferrarese} {et~al.}(2003){Ferrarese}, {C{\^o}t{\'e}}, \&
  {Jord{\'a}n}}]{fcj03}
{Ferrarese}, L., {C{\^o}t{\'e}}, P., \& {Jord{\'a}n}, A. 2003, \apj, 599, 1302

\bibitem[{{Gal-Yam} \& {Quimby}(2007)}]{gr07}
{Gal-Yam}, A. \& {Quimby}, R. 2007, The Astronomer's Telegram, 1236, 1

\bibitem[{{Henze} {et~al.}(2008){Henze}, {Pietsch}, {Burwitz},
  {Hatzidimitriou}, {Reig}, {Primak}, \& {Papamastorakis}}]{hpb+08}
{Henze}, M., {Pietsch}, W., {Burwitz}, V., {Hatzidimitriou}, D., {Reig}, P.,
  {Primak}, N., \& {Papamastorakis}, G. 2008, The Astronomer's Telegram, 1790,
  1

\bibitem[{{Hill} {et~al.}(1998){Hill}, {Nicklas}, {MacQueen}, {Mitsch},
  {Wellem}, {Altmann}, {Wesley}, \& {Ray}}]{hnm+98}
{Hill}, G.~J., {Nicklas}, H.~E., {MacQueen}, P.~J., {Mitsch}, W., {Wellem}, W.,
  {Altmann}, W., {Wesley}, G.~L., \& {Ray}, F.~B. 1998, in Society of
  Photo-Optical Instrumentation Engineers (SPIE) Conference Series, Vol. 3355,
  Society of Photo-Optical Instrumentation Engineers (SPIE) Conference Series,
  ed. {S.~D'Odorico}, 433--443

\bibitem[{{Hook} {et~al.}(2004){Hook}, {J{\o}rgensen}, {Allington-Smith},
  {Davies}, {Metcalfe}, {Murowinski}, \& {Crampton}}]{hja+04}
{Hook}, I.~M., {J{\o}rgensen}, I., {Allington-Smith}, J.~R., {Davies}, R.~L.,
  {Metcalfe}, N., {Murowinski}, R.~G., \& {Crampton}, D. 2004, \pasp, 116, 425

\bibitem[{{Hornoch} {et~al.}(2008){Hornoch}, {Scheirich}, {Garnavich},
  {Hameed}, \& {Thilker}}]{hsg+08}
{Hornoch}, K., {Scheirich}, P., {Garnavich}, P.~M., {Hameed}, S., \& {Thilker},
  D.~A. 2008, \aap, 492, 301

\bibitem[{{Hubble}(1929)}]{h29}
{Hubble}, E.~P. 1929, \apj, 69, 103

\bibitem[{{Jordi} {et~al.}(2006){Jordi}, {Grebel}, \& {Ammon}}]{jga06}
{Jordi}, K., {Grebel}, E.~K., \& {Ammon}, K. 2006, \aap, 460, 339

\bibitem[{{Kasliwal} {et~al.}(2009{\natexlab{a}}){Kasliwal}, {Cenko}, {Ofek},
  {Quimby}, {Rau}, \& {Caltech}}]{k+09c}
{Kasliwal}, M.~M., {Cenko}, S.~B., {Ofek}, E.~O., {Quimby}, R., {Rau}, A., \&
  {Caltech}, S.~R., K. 2009{\natexlab{a}}, The Astronomer's Telegram, 1984, 1

\bibitem[{{Kasliwal} {et~al.}(2009{\natexlab{b}}){Kasliwal}, {Cenko}, {Ofek},
  {Quimby}, {Rau}, \& {Kulkarni}}]{k+09b}
{Kasliwal}, M.~M., {Cenko}, S.~B., {Ofek}, E.~O., {Quimby}, R., {Rau}, A., \&
  {Kulkarni}, S.~R. 2009{\natexlab{b}}, The Astronomer's Telegram, 1934, 1

\bibitem[{{Kasliwal} {et~al.}(2008{\natexlab{a}}){Kasliwal}, {Cenko}, {Quimby},
  {Rau}, {Ofek}, \& {Kulkarni}}]{k+08f}
{Kasliwal}, M.~M., {Cenko}, S.~B., {Quimby}, R., {Rau}, A., {Ofek}, E.~O., \&
  {Kulkarni}, S.~R. 2008{\natexlab{a}}, The Astronomer's Telegram, 1880, 1

\bibitem[{{Kasliwal} {et~al.}(2008{\natexlab{b}}){Kasliwal}, {Cenko}, {Rau},
  {Ofek}, {Quimby}, \& {Kulkarni}}]{k+08e}
{Kasliwal}, M.~M., {Cenko}, S.~B., {Rau}, A., {Ofek}, E.~O., {Quimby}, R., \&
  {Kulkarni}, S.~R. 2008{\natexlab{b}}, The Astronomer's Telegram, 1854, 1

\bibitem[{{Kasliwal} {et~al.}(2008{\natexlab{c}}){Kasliwal}, {Cenko}, {Rau},
  {Ofek}, {Quimby}, \& {Kulkarni}}]{k+08b}
---. 2008{\natexlab{c}}, The Astronomer's Telegram, 1663, 1

\bibitem[{{Kasliwal} {et~al.}(2008{\natexlab{d}}){Kasliwal}, {Cenko}, {Rau},
  {Ofek}, {Quimby}, \& {Kulkarni}}]{k+08i}
---. 2008{\natexlab{d}}, The Astronomer's Telegram, 1627, 1

\bibitem[{{Kasliwal} {et~al.}(2008{\natexlab{e}}){Kasliwal}, {Cenko}, {Rau},
  {Ofek}, {Quimby}, \& {Kulkarni}}]{k+08g}
---. 2008{\natexlab{e}}, The Astronomer's Telegram, 1717, 1

\bibitem[{{Kasliwal} {et~al.}(2008{\natexlab{f}}){Kasliwal}, {Cenko}, {Rau},
  {Ofek}, {Quimby}, \& {Kulkarni}}]{k+08h}
---. 2008{\natexlab{f}}, The Astronomer's Telegram, 1719, 1

\bibitem[{{Kasliwal} {et~al.}(2008{\natexlab{g}}){Kasliwal}, {Cenko}, {Rau},
  {Ofek}, {Quimby}, \& {Kulkarni}}]{k+08d}
---. 2008{\natexlab{g}}, The Astronomer's Telegram, 1826, 1

\bibitem[{{Kasliwal} {et~al.}(2008{\natexlab{h}}){Kasliwal}, {Ofek}, {Rau},
  {Cenko}, {Quimby}, \& {Kulkarni}}]{k+08a}
{Kasliwal}, M.~M., {Ofek}, E.~O., {Rau}, A., {Cenko}, S.~B., {Quimby}, R., \&
  {Kulkarni}, S.~R. 2008{\natexlab{h}}, The Astronomer's Telegram, 1646, 1

\bibitem[{{Kasliwal} {et~al.}(2008{\natexlab{i}}){Kasliwal}, {Quimby}, {Cenko},
  {Rau}, {Ofek}, \& {Kulkarni}}]{k+08c}
{Kasliwal}, M.~M., {Quimby}, R., {Cenko}, S.~B., {Rau}, A., {Ofek}, E.~O., \&
  {Kulkarni}, S.~R. 2008{\natexlab{i}}, The Astronomer's Telegram, 1764, 1

\bibitem[{{Kasliwal} {et~al.}(2007){Kasliwal}, {Quimby}, {Rau}, {Ofek},
  {Cenko}, \& {Kulkarni}}]{k+07}
{Kasliwal}, M.~M., {Quimby}, R., {Rau}, A., {Ofek}, E., {Cenko}, B., \&
  {Kulkarni}, S. 2007, The Astronomer's Telegram, 1330, 1

\bibitem[{{Kasliwal} {et~al.}(2009{\natexlab{c}}){Kasliwal}, {Rau}, {Salvato},
  {Cenko}, {Ofek}, {Quimby}, \& {Kulkarni}}]{k+09a}
{Kasliwal}, M.~M., {Rau}, A., {Salvato}, M., {Cenko}, S.~B., {Ofek}, E.~O.,
  {Quimby}, R., \& {Kulkarni}, S.~R. 2009{\natexlab{c}}, The Astronomer's
  Telegram, 1886, 1

\bibitem[{{Kogure}(1961)}]{K61}
{Kogure}, T. 1961, \pasj, 13, 335

\bibitem[{{Kraus} \& {Hillenbrand}(2007)}]{kh07}
{Kraus}, A.~L. \& {Hillenbrand}, L.~A. 2007, \aj, 134, 2340

\bibitem[{{Law} {et~al.}(2009){Law}, {Kulkarni}, {Dekany}, {Ofek}, {Quimby},
  {Nugent}, {Surace}, {Grillmair}, {Bloom}, {Kasliwal}, {Bildsten}, {Brown},
  {Cenko}, {Ciardi}, {Croner}, {Djorgovski}, {van Eyken}, {Filippenko}, {Fox},
  {Gal-Yam}, {Hale}, {Hamam}, {Helou}, {Henning}, {Howell}, {Jacobsen},
  {Laher}, {Mattingly}, {McKenna}, {Pickles}, {Poznanski}, {Rahmer}, {Rau},
  {Rosing}, {Shara}, {Smith}, {Starr}, {Sullivan}, {Velur}, {Walters}, \&
  {Zolkower}}]{lkd+09}
{Law}, N.~M., {Kulkarni}, S.~R., {Dekany}, R.~G., {Ofek}, E.~O., {Quimby},
  R.~M., {Nugent}, P.~E., {Surace}, J., {Grillmair}, C.~C., {Bloom}, J.~S.,
  {Kasliwal}, M.~M., {Bildsten}, L., {Brown}, T., {Cenko}, S.~B., {Ciardi}, D.,
  {Croner}, E., {Djorgovski}, S.~G., {van Eyken}, J., {Filippenko}, A.~V.,
  {Fox}, D.~B., {Gal-Yam}, A., {Hale}, D., {Hamam}, N., {Helou}, G., {Henning},
  J., {Howell}, D.~A., {Jacobsen}, J., {Laher}, R., {Mattingly}, S., {McKenna},
  D., {Pickles}, A., {Poznanski}, D., {Rahmer}, G., {Rau}, A., {Rosing}, W.,
  {Shara}, M., {Smith}, R., {Starr}, D., {Sullivan}, M., {Velur}, V.,
  {Walters}, R., \& {Zolkower}, J. 2009, \pasp, 121, 1395

\bibitem[{{Lee} {et~al.}(2007){Lee}, {Ries}, {Riffeser}, \& {Seitz}}]{lrr+07}
{Lee}, C., {Ries}, C., {Riffeser}, A., \& {Seitz}, S. 2007, The Astronomer's
  Telegram, 1324, 1

\bibitem[{{Livio}(1992)}]{l92}
{Livio}, M. 1992, \apj, 393, 516

\bibitem[{{McLaughlin}(1960)}]{m60}
{McLaughlin}, D.~B. 1960, in Stellar Atmospheres, ed. {J.~L.~Greenstein},
  585--+

\bibitem[{{Oke} {et~al.}(1995){Oke}, {Cohen}, {Carr}, {Cromer}, {Dingizian},
  {Harris}, {Labrecque}, {Lucinio}, {Schaal}, {Epps}, \& {Miller}}]{occ+95}
{Oke}, J.~B., {Cohen}, J.~G., {Carr}, M., {Cromer}, J., {Dingizian}, A.,
  {Harris}, F.~H., {Labrecque}, S., {Lucinio}, R., {Schaal}, W., {Epps}, H., \&
  {Miller}, J. 1995, \pasp, 107, 375

\bibitem[{{Oke} \& {Gunn}(1982)}]{og82}
{Oke}, J.~B. \& {Gunn}, J.~E. 1982, \pasp, 94, 586

\bibitem[{{Ovcharov} {et~al.}(2007){Ovcharov}, {Valcheva}, {Kostov}, {Nikolov},
  {Georgiev}, \& {Nedialkov}}]{ovk+07}
{Ovcharov}, E., {Valcheva}, A., {Kostov}, A., {Nikolov}, Y., {Georgiev}, T., \&
  {Nedialkov}, P. 2007, The Astronomer's Telegram, 1312, 1

\bibitem[{{Payne-Gaposchkin}(1964)}]{p64}
{Payne-Gaposchkin}, C. 1964, {The galactic novae}, ed. {Gaposchkin, C.~H.~P.}

\bibitem[{{Pietsch} {et~al.}(2007){Pietsch}, {Burwitz}, {Stoss}, {Updike},
  {Hartmann}, {Milne}, \& {Williams}}]{pbs+07}
{Pietsch}, W., {Burwitz}, V., {Stoss}, R., {Updike}, A., {Hartmann}, D.,
  {Milne}, P., \& {Williams}, G. 2007, The Astronomer's Telegram, 1230, 1

\bibitem[{{Rahmer} {et~al.}(2008){Rahmer}, {Smith}, {Velur}, {Hale}, {Law},
  {Bui}, {Petrie}, \& {Dekany}}]{gsv+08}
{Rahmer}, G., {Smith}, R., {Velur}, V., {Hale}, D., {Law}, N., {Bui}, K.,
  {Petrie}, H., \& {Dekany}, R. 2008, in Presented at the Society of
  Photo-Optical Instrumentation Engineers (SPIE) Conference, Vol. 7014, Society
  of Photo-Optical Instrumentation Engineers (SPIE) Conference Series

\bibitem[{{Rau} {et~al.}(2009{\natexlab{a}}){Rau}, {Kasliwal}, \&
  {Salvato}}]{rks09}
{Rau}, A., {Kasliwal}, M.~M., \& {Salvato}, M. 2009{\natexlab{a}}, The
  Astronomer's Telegram, 1887, 1

\bibitem[{{Rau} {et~al.}(2009{\natexlab{b}}){Rau}, {Kulkarni}, {Law}, {Bloom},
  {Ciardi}, {Djorgovski}, {Fox}, {Gal-Yam}, {Grillmair}, {Kasliwal}, {Nugent},
  {Ofek}, {Quimby}, {Reach}, {Shara}, {Bildsten}, {Cenko}, {Drake},
  {Filippenko}, {Helfand}, {Helou}, {Howell}, {Poznanski}, \&
  {Sullivan}}]{rkl+09}
{Rau}, A., {Kulkarni}, S.~R., {Law}, N.~M., {Bloom}, J.~S., {Ciardi}, D.,
  {Djorgovski}, G.~S., {Fox}, D.~B., {Gal-Yam}, A., {Grillmair}, C.~C.,
  {Kasliwal}, M.~M., {Nugent}, P.~E., {Ofek}, E.~O., {Quimby}, R.~M., {Reach},
  W.~T., {Shara}, M., {Bildsten}, L., {Cenko}, S.~B., {Drake}, A.~J.,
  {Filippenko}, A.~V., {Helfand}, D.~J., {Helou}, G., {Howell}, D.~A.,
  {Poznanski}, D., \& {Sullivan}, M. 2009{\natexlab{b}}, \pasp, 121, 1334

\bibitem[{{Schaefer}(2009)}]{s09}
{Schaefer}, B.~E. 2009, ArXiv e-prints

\bibitem[{{Shafter} {et~al.}(2008){Shafter}, {Ciardullo}, {Bode}, {Darnley}, \&
  {Misselt}}]{scb+08}
{Shafter}, A.~W., {Ciardullo}, R., {Bode}, M.~F., {Darnley}, M.~J., \&
  {Misselt}, K.~A. 2008, The Astronomer's Telegram, 1832, 1

\bibitem[{{Shafter} {et~al.}(2009){Shafter}, {Rau}, {Quimby}, {Kasliwal},
  {Bode}, {Darnley}, \& {Misselt}}]{srq+09}
{Shafter}, A.~W., {Rau}, A., {Quimby}, R.~M., {Kasliwal}, M.~M., {Bode}, M.~F.,
  {Darnley}, M.~J., \& {Misselt}, K.~A. 2009, \apj, 690, 1148

\bibitem[{{Shara}(1981)}]{s81}
{Shara}, M.~M. 1981, \apj, 243, 926

\bibitem[{{Shen} \& {Bildsten}(2009)}]{sb09}
{Shen}, K.~J. \& {Bildsten}, L. 2009, \apj, 692, 324

\bibitem[{{Townsley} \& {Bildsten}(2004)}]{tb04}
{Townsley}, D.~M. \& {Bildsten}, L. 2004, \apj, 600, 390

\bibitem[{{Valcheva} {et~al.}(2008){Valcheva}, {Ovcharov}, {Latev}, {Kostov},
  {Nikolov}, {Georgiev}, \& {Nedialkov}}]{vol+08}
{Valcheva}, A., {Ovcharov}, E., {Latev}, G., {Kostov}, A., {Nikolov}, Y.,
  {Georgiev}, T., \& {Nedialkov}, P. 2008, The Astronomer's Telegram, 1687, 1

\bibitem[{{van den Bergh} \& {Younger}(1987)}]{vy87}
{van den Bergh}, S. \& {Younger}, P.~F. 1987, \aaps, 70, 125

\bibitem[{{Williams}(1992)}]{w92}
{Williams}, R.~E. 1992, \aj, 104, 725

\bibitem[{{Williams} {et~al.}(1994){Williams}, {Phillips}, \& {Hamuy}}]{wph94}
{Williams}, R.~E., {Phillips}, M.~M., \& {Hamuy}, M. 1994, \apjs, 90, 297

\bibitem[{{Yaron} {et~al.}(2005){Yaron}, {Prialnik}, {Shara}, \&
  {Kovetz}}]{yps+05}
{Yaron}, O., {Prialnik}, D., {Shara}, M.~M., \& {Kovetz}, A. 2005, \apj, 623,
  398

\bibitem[{{Zwicky}(1936)}]{z36}
{Zwicky}, F. 1936, \pasp, 48, 191

\end{thebibliography}

\end{document}